\documentclass[reprint,twocolumn,floatfix,aip,jcp,graphicx]{revtex4-2}
\draft 

\usepackage{placeins}
\usepackage{tabularx}
\usepackage{amsmath}
\usepackage{amsfonts}
\usepackage{amssymb}
\usepackage{graphicx}
\usepackage{chemformula}
\usepackage{float}
\usepackage[version=4]{mhchem}
\usepackage{threeparttable}
\usepackage[dvipsnames]{xcolor}
\usepackage[shortlabels]{enumitem}
\usepackage{dcolumn} 
\usepackage{tikz}
\usetikzlibrary{shapes.geometric, arrows.meta, positioning}
\newcolumntype{d}[1]{D{.}{.}{#1}}
\newcommand{\ket}[1]{\ensuremath{\left| \left. {#1} \right\rangle \right.}}

\usepackage{todonotes}

\definecolor{darkgreen}{HTML}{009933}

\begin{document}

\title[]{Basis pruning in rovibrational configuration interaction (RVCI) calculations}

\author{Kemal Oenen}
\affiliation{Institute for Theoretical Chemistry, University of Stuttgart, Pfaffenwaldring 55, 70569 Stuttgart, Germany}%
\affiliation{Institute for Quantum Electronics, Department of Physics, ETH Zürich, Otto-Stern-Weg 1, 8093 Zürich, Switzerland}

\author{Subhasish Das}
\affiliation{Institute for Theoretical Chemistry, University of Stuttgart, Pfaffenwaldring 55, 70569 Stuttgart, Germany}%

\author{Guntram Rauhut$^{*,}$}
\email[Author to whom correspondence should be addressed: ]{rauhut@theochem.uni-stuttgart.de}
\affiliation{Institute for Theoretical Chemistry, University of Stuttgart, Pfaffenwaldring 55, 70569 Stuttgart, Germany}%

\date{\today}

\begin{abstract}
The variational calculation of rovibrational spectra without model Hamiltonians remains challenging, with the size and composition of the vibrational basis constituting a major bottleneck. Preceding vibrational configuration interaction (VCI) calculations reduce this basis, but leave the question of which vibrational state functions to include. In close analogy to configuration-selective VCI theory, we propose a basis-pruning algorithm in which candidate vibrational states are represented by inexpensive approximate wavefunctions and selected according to their impact on rovibrational energies in a spectral window of interest. For formaldehyde, comparisons with widely used polyad schemes controlling the vibrational basis are provided. Calculations on thioketene benchmark the accuracy and efficiency of the pruning approach for different threshold values.
\end{abstract}

\maketitle

\section{Introduction}\label{sec:introduction}
Computational high-resolution molecular spectroscopy by means of variational rovibrational calculations becomes increasingly demanding with molecular size, limiting its practical application primarily to small molecules \cite{RVCI_book_chapter, YurchenkoBook2024, Avila2019, Carrington2018Review}. This limitation arises, \textit{inter alia}, from the need for a multidimensional Born--Oppenheimer potential energy surface (PES) of very high accuracy and from the rapidly growing size of the vibrational basis entering the subsequent rovibrational treatment \cite{Erfort2020, Tschoepe2022, Matyus2023}. With respect to the former, modern computer architectures in combination with efficient explicitly correlated coupled-cluster methods \cite{ccf12}, an $n$-mode expansion of the PES \cite{n-mode}, and the use of well-balanced multi-level schemes \cite{MultiLevel,ml_yagi,ml_ove,Schroeder2024} allow for the calculation of highly accurate PESs. For systems with up to 10 or 12 atoms, such PESs can reproduce experimental gas-phase fundamental frequencies \cite{Rauhut2004,Ziegler2016,Ziegler2018,Schroeder2024} with mean absolute deviations of 2--3~cm$^{-1}$. 

Considering the vibrational wavefunction in terms of Hartree products, one may distinguish between approaches in which these functions enter the rovibrational calculation directly and those in which a preceding purely vibrational calculation is used to generate a contracted basis of vibrational state functions (VSFs), i.e., linear combinations of Hartree products. Here, we adopt the latter strategy within the framework of rovibrational configuration interaction (RVCI) theory. Such a basis can, for example, be generated by configuration-selective VCI (cs-VCI) calculations \cite{Schroeder2022, Neff2009, Rauhut2007, Mathea2021, Mathea2022} that have been applied to systems with up to 20 atoms \cite{Oschetzki2014}. While this substantially reduces the basis size compared with a direct Hartree-product ansatz, the resulting set of VSFs may nevertheless remain large, even though only a subset is typically required for converged rovibrational calculations \cite{RVCI_book_chapter, Tschoepe2022}. This raises the question: which ones should be selected?

The principle of retaining only those basis functions needed to maintain sufficient accuracy is commonly referred to as basis pruning and has long been employed in theoretical molecular spectroscopy \cite{Carter1986, Carrington2018Review, Matyus2023}. Simple choices are restrictions on the sum of the vibrational quantum numbers and/or on the energy of the basis functions \cite{YurchenkoBook2024, RVCI_book_chapter}. These criteria can be applied both to prune the Hartree-product basis underlying a vibrational calculation \cite{Halonen1983, WangCarrington2001, AvilaCarrington2011, Wang2015, Avila2019, Alberto2022, Simmons2023, Sunaga2025} and to control which of the correlated VSFs obtained in this way are included in the rovibrational basis \cite{YurchenkoBook2024, Tschoepe2021, Erfort2022, Schneider2022, Tschoepe2023a, Tschoepe2023b, DeVos2023, Das2024b, Das2024c, Das2024d, Oenen2025}. \\ \indent More physically motivated pruning conditions can be constructed from weighted sums of vibrational quantum numbers \cite{AvilaCarrington2009}, with polyad schemes providing a prominent example \cite{Polyad_Review, Rey2022NovelAnalysis, Krasnoshchekov2013, Kellman1995, Kellman1990, Amrein1985, Ishikawa1998, Barnes2011, BermudezMontana2017, Nikitin2024, Rodina2026}. Here, integer weights are chosen to reflect ratios among the harmonic frequencies, such that states sharing the same value of the sum, termed the polyad number, approximately cluster in energy. The basis can then be pruned by imposing an upper cutoff on this number \cite{YurchenkoBook2024, Yachmenev2011, Yachmenev2013, SousaSilva2013, Refaie015}. The usefulness of such schemes, however, depends on identifying an appropriate polyad structure, which becomes increasingly difficult for larger, less symmetric molecules \cite{RVCI_book_chapter, YurchenkoBook2024, Rey2022NovelAnalysis, Krasnoshchekov2013}. Beyond linear weighted sums, more general pruning functions can distinguish states with excitation distributed over many modes from those with higher excitation confined to only a few modes \cite{Avila2012, Avila2017}. \\ \indent Importantly, these approaches allow the set of correlated VSFs to be selected in advance, so that only those entering the rovibrational problem need to be computed. Their cutoff parameters, however, generally have to be established empirically through increasingly demanding convergence calculations \cite{Yachmenev2013, SousaSilva2013, Tschoepe2022, Sunaga2025, Avila2019, Alberto2022}, while the underlying criteria do not directly indicate whether a particular state is important for the spectral region of interest. Consequently, unnecessary states may be retained while influential ones can still be omitted. Once the correlated eigenstates are available, alternative pruning strategies based on properties such as vibrational transition intensities can also be employed \cite{Owens2017, Mant2018, Yurchenko2024}. Nevertheless, it is generally desirable to avoid the costly computation of correlated wavefunctions that are ultimately discarded. \\ \indent A different perspective emerges when basis functions are judged by the effect they exert on the states one seeks to describe. In electronic structure theory (EST), selected-CI methods have long followed this idea by drawing only those configurations from the vast configuration space whose estimated importance for the target CI solution is sufficiently large \cite{Whitten1969, Huron1973, Buenker1974, Stampfuss1999, Schriber_JCTC13_2017}. A closely related strategy is found in vibrational structure theory (VST), notably in cs-VCI, where Hartree-product configurations are tested iteratively and admitted according to perturbative estimates of their contribution to a target VCI state \cite{Mathea2022, Schroeder2022, Rauhut2007, Scribano2008, Carbonniere_TCA125_2010, Handy2004, Wang2015}. This suggests applying the same rationale directly to the pruning of the vibrational basis used in RVCI. In doing so, the relevance of a candidate VSF must be assessed by its collective impact on rovibrational energies across a spectral window, rather than by its effect on a single level. Inevitably, however, a literal transfer creates a circular dependency, since, unlike the inexpensive configurations screened in cs-VCI, the VSFs required for such an assessment would first have to be computed. Generating them merely to determine whether they are needed would thus effectively defeat the purpose of pruning. \\ \indent To circumvent this circular dependency, we represent a broad set of candidate VSFs by second-order vibrational M{\o}ller–Plesset (VMP2)-like wavefunctions \cite{christiansen2003, Mathea2022}, which can be computed efficiently in reduced configuration spaces. The candidates are evaluated by the changes they induce in rovibrational energies within a user-defined spectral region and retained when these exceed a chosen screening threshold. Accurate VCI wavefunctions are subsequently computed only for the resulting subset and used in the final RVCI calculations. We have implemented this basis-pruning scheme in the \textsc{Molpro} package of \textit{ab initio} programs \cite{Werner2020} and demonstrate its performance for formaldehyde through comparison with polyad-based truncations and for thioketene over a range of screening thresholds.

\section{Theory}\label{sec:theory}

This section introduces the theoretical framework and notation     adopted throughout this work. We first briefly outline the RVCI formalism (cf.\ Section~\ref{subsec:rvci_theory}). Next, we discuss the concepts underlying vibrational basis pruning and introduce the auxiliary state functions employed for this purpose (cf.\ Section~\ref{subsec:pruning_theory}). Finally, we define the pruning metric (cf.\ Section~\ref{subsec:pruning_metrics}).

\subsection{Rovibrational configuration interaction (RVCI) theory}\label{subsec:rvci_theory}
The RVCI formalism\cite{RVCI_book_chapter,Erfort2020,Tschoepe2022}, as available in \textsc{Molpro}\cite{Werner2020}, is based on the Watson Hamiltonian\cite{Watson1968} for non-linear polyatomic molecules. It can be decomposed into rotational ($\hat{\mathcal{H}}_{\mathrm{rot}}$), Coriolis-coupling ($\hat{\mathcal{H}}_{\mathrm{cor}}$), and vibrational ($\hat{\mathcal{H}}_{\mathrm{vib}}$) contributions,
\begin{equation}\label{eq:watsonian}
    \hat{\mathcal{H}}_{\mathrm{Watson}}
    =
    \hat{\mathcal{H}}_{\mathrm{rot}}
    +
    \hat{\mathcal{H}}_{\mathrm{cor}}
    +
    \hat{\mathcal{H}}_{\mathrm{vib}}
\end{equation}
with
\begin{equation}\label{eq:watsonian_terms}
\begin{aligned}
    \hat{\mathcal{H}}_{\mathrm{rot}}
    &= \frac{1}{2}\sum_{\alpha\beta} \hat{J}_\alpha\,\mu_{\alpha\beta}\,\hat{J}_\beta,\\
    \hat{\mathcal{H}}_{\mathrm{cor}}
    &= -\frac{1}{2}\sum_{\alpha\beta}\!\left(
    \hat{J}_\alpha\,\mu_{\alpha\beta}\,\hat{\pi}_\beta
    +\hat{\pi}_\alpha\,\mu_{\alpha\beta}\,\hat{J}_\beta\right),\\
    \hat{\mathcal{H}}_{\mathrm{vib}}
    &= \frac{1}{2}\sum_{\alpha\beta}\hat{\pi}_\alpha\,\mu_{\alpha\beta}\,\hat{\pi}_\beta
    -\frac{1}{2}\sum_i \frac{\partial^2}{\partial q_i^2}
    \\ &-\frac{1}{8}\sum_\alpha \mu_{\alpha\alpha}
    +V(\mathbf{q}).
\end{aligned}
\end{equation}
Here, the indices $\alpha,\beta\in\{x,y,z\}$ label the Cartesian coordinates, and $\hat{J}_\alpha$ denotes the corresponding component of the angular momentum operator. For an $N$-atomic system, the potential energy $V(\mathbf{q})$ is expressed in terms of an $n$-mode expansion\cite{Ziegler2016, Ziegler2018} with respect to canonical normal coordinates $\mathbf{q}=(q_1,\dots,q_{3N-6})$. The vibrational angular momentum (VAM) operator $\hat{\pi}_\alpha$ entering the Coriolis-coupling and vibrational terms is given by 
\begin{equation}\label{eq:vam_operator}
    \hat{\pi}_\alpha = -i\sum_{ij}^{3N-6}\zeta^{\alpha}_{ij}\,q_i\,\frac{\partial}{\partial q_j},
\end{equation}
where $\zeta^{\alpha}_{ij}$ are the elements of the antisymmetric Coriolis $\zeta$-coupling matrices, which depend on the displacement vectors of the normal coordinates. The effective moment of inertia tensor
\begin{equation}\label{eq:eff_mom_of_intertia_ensor}
    I'_{\alpha\beta}(\mathbf{q}) = I_{\alpha\beta}(\mathbf{q})-\sum_{ijk}\zeta^{\alpha}_{ij}\zeta^{\beta}_{kj}\,q_i q_k,
\end{equation}
determines the $\mu$-tensor through the relation
\begin{equation}\label{eq:mu_tensor_def}
    \boldsymbol{\mu}(\mathbf{q})=\big(\mathbf{I}'(\mathbf{q})\big)^{-1}.
\end{equation}
Mirroring the treatment of $V(\mathbf{q})$, $\boldsymbol{\mu}(\mathbf{q})$ can also be represented in terms of an $n$-mode expansion and fitted to a polynomial basis\cite{Neff2011},
\begin{equation}\label{eq:mu_expansion}
    \mu_{\alpha\beta}(\mathbf{q}) = \mu_{\alpha\beta}^{(0)} + \sum_{i}\sum_{r} \mu_{\alpha\beta,i}^{(1),r} q_i^r + \sum_{i<j}\sum_{rs} \mu_{\alpha\beta,ij}^{(2),rs} q_i^r q_j^s + \dots .
\end{equation}
Note that, in this form, the Coriolis-coupling contribution cannot, in general, be simplified further because the VAM operator does not commute with the individual terms of the $\boldsymbol{\mu}$-tensor expansion.

Within the variational framework, the matrix representation of $\hat{\mathcal{H}}_{\mathrm{Watson}}$ is obtained by expanding the rovibrational wavefunction, $\ket{\Psi_{\mathrm{rovib}}}$, in a direct-product basis of rotational and vibrational functions\cite{Erfort2020, Tschoepe2022},
\begin{equation}\label{eq:rvci_ansatz}
    \ket{\Psi_{\mathrm{rovib}}}
    =
    \sum_{r\nu}
    c_{r\nu}
    \ket{\Phi_r}
    \ket{\Phi_\nu},
\end{equation}
where the rotational basis $\{\ket{\Phi_r}\}$ is chosen as Wang combinations\cite{Spirko1985,Yurchenko2005,Das2024}.

To construct $\{\ket{\Phi_\nu}\}$, we solve the Schrödinger equation for the Hamiltonian $\hat{\mathcal{H}}_{\mathrm{vib}}$ using VCI theory\cite{Neff2009, Schroeder2022}. The resulting symmetry-adapted states form the contracted vibrational basis for the subsequent RVCI calculation. Each state is expanded as a linear combination of configurations $\{\ket{\Psi_I}\}$,
\begin{equation}\label{eq:vib_vci_expansion}
    \ket{\Phi_\nu}
    =
    \sum_I c_\nu^{I}\ket{\Psi_I}
    =
    \sum_I c_\nu^{I}\prod_i \varphi_i^I(q_i),
\end{equation}
where the coefficients $c_\nu^{I}$ are determined variationally. The one-mode wavefunctions (modals) $\varphi_i^I$ are obtained from a preceding vibrational self-consistent field (VSCF) calculation \cite{Bowman1986, Carter1997, Rauhut2004} and expanded in a primitive basis of distributed Gaussians \cite{Hamilton1986}. In practice, configuration-selection schemes \cite{Rauhut2007, Mathea2022, Schroeder2022} are used to limit the exponentially growing correlation space by retaining only those Hartree products that contribute significantly to the vibrational states of interest (cf.\ Section~\ref{subsec:pruning_theory}).

For a given rotational quantum number $J$, the rotational basis comprises at most $2J+1$ meaningful functions, whereas no analogous upper bound applies to the vibrational basis. While a sufficiently large vibrational basis is required to obtain reliable rovibrational energies, increasing its size $N_{\mathrm{vib}}$ rapidly raises the computational effort, since the dimension of the Hamiltonian scales as $N_{\mathrm{vib}}(2J+1)$. Consequently, converging the vibrational basis is particularly demanding because $N_{\mathrm{vib}}$ affects essentially all time-consuming stages of both the VCI and RVCI calculations through the dimensions of the underlying eigenvalue problems. By comparison, other convergence parameters, such as the truncation of the $\boldsymbol{\mu}$-tensor expansion, typically have a more limited influence on the overall computational cost. The basis size alone, however, does not determine which vibrational states should be included. Their relevance depends on the molecule and on the particular rovibrational energies targeted within a given spectral region, making the selection of an appropriate set an additional challenge \cite{Tschoepe2022}.

With the direct-product basis of Eq.\ \eqref{eq:rvci_ansatz}, the Hamiltonian matrix elements associated with the individual terms in Eq.\ \eqref{eq:watsonian_terms} can be written as products of rotational and vibrational integrals\cite{Tschoepe2022}. Specifically, the vibrational contributions entering $\hat{\mathcal{H}}_{\mathrm{rot}}$ and $\hat{\mathcal{H}}_{\mathrm{cor}}$ are of the form
\begin{equation}\label{eq:vib_integrals}
    \langle v'|\mu_{\alpha\beta}|v''\rangle
    \quad \text{and} \quad
    \langle v'|\mu_{\alpha\beta}\hat{\pi}_\beta|v''\rangle,
\end{equation}
respectively. Since these vibrational integrals are independent of $J$, they can be precomputed once and reused for the sequence of rovibrational matrices required in an RVCI calculation. Notably, although the discussion above is phrased in terms of VCI state functions, the formalism is general and only requires a basis for which the classes of integrals in Eq.\ \eqref{eq:vib_integrals} can be evaluated. This observation motivates the auxiliary state functions introduced in Section~\ref{subsec:pruning_theory}. Further details on RVCI theory and its \textsc{Molpro} implementation, such as the calculation of transition intensities, are given in Refs.\ \citenum{Erfort2020, Tschoepe2022, RVCI_book_chapter}. 

\subsection{Vibrational basis pruning for RVCI}\label{subsec:pruning_theory}

In configuration-selective VCI (cs-VCI), the correlation space, i.e., the set of configurations $\{\ket{\Psi_I}\}$ appearing in the VCI expansion of Eq.\ \eqref{eq:vib_vci_expansion}, is constructed iteratively for each target state \cite{Mathea2022}. In iteration $n$, an intermediate VCI diagonalization within the current space $\mathcal{C}_\nu^{(n)}$ yields the approximate wave function
\begin{equation}
    \ket{\Phi_\nu^{(n)}}=
    \sum_{I\in\mathcal{C}_\nu^{(n)}}
    c_{\nu I}^{(n)}\ket{\Psi_I}.
\end{equation}
The influence of an external candidate configuration $\ket{\Psi_K}$ is estimated using a VMP2-like selection criterion\cite{Rauhut2007,Mathea2022},
\begin{equation}\label{eq:vmp2_importance}
    \varepsilon_{\nu K}^{(n+1)}
    =
    \frac{
    \big|
    \langle\Phi_\nu^{(n)}|\hat V|\Psi_K\rangle
    \big|^2
    }{
    \varepsilon_\nu^{(n)}-\varepsilon_K
    },
\end{equation}
where $\hat V$ denotes the perturbation operator, $\varepsilon_\nu^{(n)}$ is the energy of the current approximate state, and $\varepsilon_K$ the energy of $\ket{\Psi_K}$. If $\varepsilon_{\nu K}^{(n+1)}$ exceeds a prescribed threshold, $\ket{\Psi_K}$ is added to $\mathcal{C}_\nu^{(n)}$. The configuration space is subsequently enlarged and rediagonalized until the target-state energy converges. This iterative procedure also accounts for configurations that become important through indirect coupling to those selected in earlier iterations. By pruning the correlation space to important configurations, cs-VCI substantially reduces the CPU time and memory requirements of VCI calculations \cite{Rauhut2007,Schroeder2022}.

For RVCI calculations in \textsc{Molpro}, the contracted vibrational basis is commonly controlled through upper bounds on the excitation level and state energy \cite{RVCI_book_chapter,Tschoepe2022}. As motivated in Section~\ref{sec:introduction}, we pursue a pruning strategy for this basis in close analogy to cs-VCI, with correlated vibrational eigenstates taking the place of individual configurations. To avoid computing a VCI wavefunction for every candidate solely for screening, we rely on approximate auxiliary VSFs. Pure VSCF functions would provide the simplest choice, but their mean-field character neglects mode correlation and may therefore inadequately describe couplings between vibrational states. We instead use a VMP2-like ansatz, which incorporates correlation beyond the mean-field level while retaining the Hartree-product expansion of the final VCI states. This representation also allows the classes of vibrational integrals in Eq.\ \eqref{eq:vib_integrals} to be evaluated in the same form, with the expansion coefficients determined perturbatively rather than variationally.

Specifically, we construct the renormalized auxiliary state functions as
\begin{equation}\label{eq:aux_state_expansion}
    \ket{\widetilde{\chi}_\nu}
    =
    \sum_{I\in\widetilde{\mathcal{C}}_\nu}
    \widetilde{c}^{I}_\nu\ket{\Psi_I},
    \qquad
    \widetilde{\mathcal{C}}_\nu\subset\mathcal{C}_\nu,
\end{equation}
where $\widetilde{\mathcal{C}}_\nu$ is chosen substantially smaller than the correlation space $\mathcal{C}_\nu$ employed in the subsequent VCI calculation. The functions are formed from ground-state modals using a fixed VSCF reference and are orthogonalized before entering the pruning procedure. They are not intended to provide quantitatively accurate energies, but rather to serve as inexpensive proxies for assessing the influence of candidate vibrational states on the rovibrational levels within a spectral window of interest. In practice, this effect is determined from the shifts induced in these levels when the auxiliary state is included in a reduced RVCI calculation, as described in Section~\ref{sec:methodology}. These shifts form the basis of the pruning metric introduced in the following section.

\subsection{Pruning metric}\label{subsec:pruning_metrics}

Let $\{E_k^{\mathrm{ref}}\}$ denote the rovibrational energies from an RVCI calculation without a vibrational test state $c$, and $\{E_k^{\mathrm{test}}(c)\}$ those from an otherwise identical calculation including $c$. After matching the rovibrational levels based on their quantum-number assignments\cite{Mathea2020,Conway2021}, the absolute energy shifts are obtained as
\begin{equation}\label{eq:deltaE_level}
    \Delta E_k(c)
    =
    \left|
    E_k^{\mathrm{test}}(c)-E_k^{\mathrm{ref}}
    \right|,
    \qquad
    k\in\mathcal{K},
\end{equation}
where $\mathcal{K}$ contains the matched levels included in the comparison (cf.\ Section~\ref{sec:methodology}). To condense these shifts into a single quantity, we take their third quartile,
\begin{equation}\label{eq:deltaE_Q3}
    Q_3(c)
    =
    \operatorname{quantile}_{0.75}
    \big(
    \{\Delta E_k(c)\}_{k\in\mathcal{K}}
    \big).
\end{equation}
This choice remains sensitive to the larger changes within the matched set while limiting the influence of isolated outliers \cite{Tschoepe2022}. The quantity $Q_3(c)$ constitutes the pruning metric, i.e., a candidate is retained whenever $Q_3(c)$ exceeds a chosen threshold. With the theoretical framework in place, we now turn to the practical implementation of the pruning approach.

\section{Methodology}\label{sec:methodology}

\begin{figure}[ht!]
    \centering
    \includegraphics[width=\columnwidth, keepaspectratio]{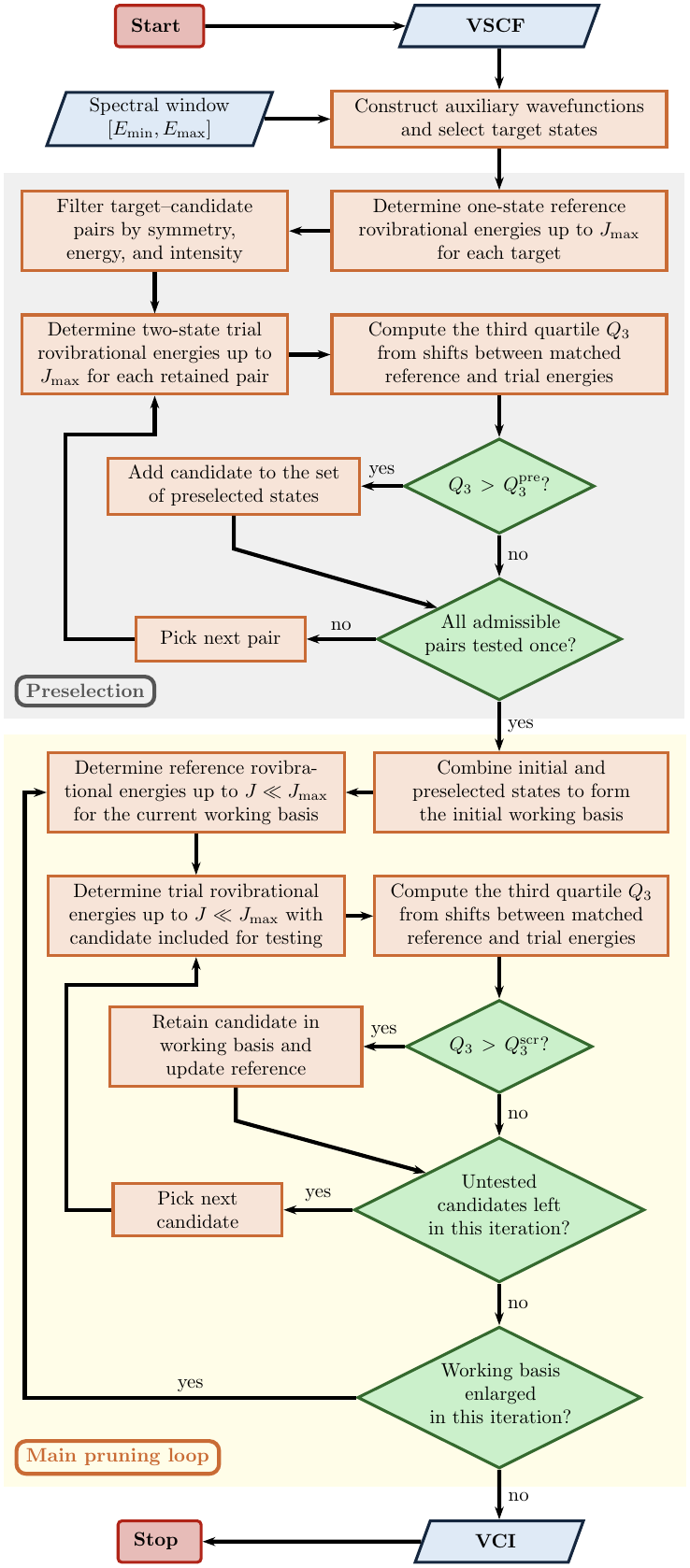}
    \caption{Overview of the vibrational basis pruning scheme implemented in \textsc{Molpro} \cite{Werner2020}. A preselection screens each admissible target--candidate pair up to $J_{\max}$ for interactions that may become relevant at higher $J$ (grey background). The main pruning loop then tests the remaining candidates iteratively at low $J$ until the working basis no longer changes (yellow background). The resulting pruned VCI basis is used in the final RVCI calculation.}
    \label{fig:rvci_pruning_flowchart}
\end{figure}

Figure~\ref{fig:rvci_pruning_flowchart} summarizes the pruning workflow employed in this work. The starting point is a VSCF calculation, which provides, \textit{inter alia}, the optimized ground-state modals defining the underlying Hartree-product basis, as well as state descriptors such as symmetry labels and intensities. The VSCF state list is characterized by the quantum-number vectors
$\mathbf n(k)=(n_1(k),\dots,n_{3N-6}(k))$, where $k$ is the state index and $n_i(k)$ denotes the number of quanta in mode $i$. Based on this list, we construct the approximate wavefunctions of Eq.~\eqref{eq:aux_state_expansion}, yielding the total state set $\mathcal{S}_{\mathrm{total}}$ together with the VMP2-like energies $E_k$. The vibrational matrix elements of Eq.~\eqref{eq:vib_integrals} are precomputed for all members of $\mathcal{S}_{\mathrm{total}}$ and reused during the pruning calculations. For the spectral region of interest, the user specifies an energy interval $[E_{\min},E_{\max}]$, from which the vibrational target states are selected as
\begin{equation}
    \begin{aligned}
        \mathcal{S}_{\mathrm{tar}}
        = \Big\{\,\ket{\tilde{\chi}_k}\in\mathcal{S}_{\mathrm{total}}
        \;\Big|\;&
        \sum_{i=1}^{3N-6} n_i(k)=1,\\
        & E_k\in[E_{\min},E_{\max}]
        \,\Big\}.
    \end{aligned}
\end{equation}
Together with the ground state, $\mathcal{S}_{\mathrm{tar}}$ forms the initial basis,
$\mathcal{S}_{\mathrm{init}}=\{\ket{\tilde{\chi}_0}\}\cup\mathcal{S}_{\mathrm{tar}}$.
This compact basis is subsequently enlarged according to the pruning criterion.

States in $\mathcal{S}_{\mathrm{total}}$ that are not contained in the initial basis are considered candidates for addition. Each candidate is tested by comparing RVCI calculations with and without that state, considering only rovibrational levels assigned to the vibrational targets defined above. For the applications presented here, only levels associated with the fundamentals in $\mathcal{S}_{\mathrm{tar}}$ are monitored, although ground-state levels may optionally be included. Carrying out this screening for every candidate up to the maximum rotational quantum number $J_{\max}$ of the final RVCI calculation would be unnecessarily demanding as the basis grows. We therefore restrict the main pruning loop to small $J$, using $J=10$ throughout this study. Interactions that become appreciable only at higher $J$ through the rotational or Coriolis-coupling terms of the Watson Hamiltonian may consequently be overlooked. In particular, omission of an important interaction partner may qualitatively alter the predicted band structure, e.g., when an avoided crossing gives rise to a band head (cf.\ Section~\ref{subsec:h2ccs}). To account for such interactions, a preselection extending to $J_{\max}$ is introduced.

For each monitored target state $\ket{\tilde{\chi}_t}$, a reduced one-state RVCI calculation is performed for all $J\leq J_{\max}$. The rovibrational levels assigned to the target are filtered using an assignment-confidence threshold of $A_{\min}=0.5$ to ensure reliable matching in the following comparison \cite{Tschoepe2022} (cf.\ Section~\ref{subsec:pruning_metrics}). These levels form the one-state reference used to assess the effect of adding a candidate. Since testing every possible target--candidate pair would require a large number of RVCI calculations, the candidate list is narrowed beforehand. Accordingly, we consider only pairs with the same vibrational symmetry, $\Gamma_c=\Gamma_t$, an energy separation $|E_c-E_t|\leq\delta_{\mathrm{thr}}$, and at least one state exceeding a VSCF intensity threshold $I_{\mathrm{thr}}$. The values $\delta_{\mathrm{thr}}=750~\mathrm{cm}^{-1}$ and $I_{\mathrm{thr}}=10^{-2}~\mathrm{km\,mol^{-1}}$ are adopted for the present work. While the symmetry and energy criteria retain a manageable subset whose rovibrational levels were found to be particularly prone to avoided crossings at higher $J$, the intensity criterion removes pairs unlikely to contribute visible spectral features. We note that these constraints are specific to the preselection, while the main pruning loop (\textit{vide infra}) does not rely on them.

A two-state trial RVCI calculation up to $J_{\max}$ is then carried out for every remaining target--candidate pair. Following the same filtering, the rovibrational levels associated with the target are matched to the one-state reference. The resulting set of matched levels corresponds to $\mathcal{K}$ in Section~\ref{subsec:pruning_metrics}, and their energy shifts are used to evaluate the pruning metric $Q_3$. Candidates with $Q_3>Q_3^{\mathrm{pre}}=10^{-1}~\mathrm{cm}^{-1}$ are added to the preselection set $\mathcal{S}_{\mathrm{presel}}$.

The main pruning loop consequently starts from the working basis
$\mathcal{S}_{\mathrm{work}}=\mathcal{S}_{\mathrm{init}}\cup\mathcal{S}_{\mathrm{presel}}$. Although this basis is considerably larger than the one- and two-state bases used in the preselection, restricting the calculations to $J\leq10$ keeps the repeated RVCI diagonalizations computationally feasible. At the beginning of each iteration, a reference calculation is performed for the current $\mathcal{S}_{\mathrm{work}}$ using the same assignment-confidence filtering as in the preselection. Subsequently, for each candidate $\ket{\tilde{\chi}_c}\notin\mathcal{S}_{\mathrm{work}}$, the enlarged basis $\mathcal{S}_{\mathrm{work}}\cup\{\ket{\tilde{\chi}_c}\}$ provides a new set of trial rovibrational eigenvalues. After matching these to the reference values, the corresponding shifts are used to compute $Q_3$ (cf.\ Section~\ref{subsec:pruning_metrics}), considering only the monitored target levels. Whenever this value is larger than the default screening threshold $Q_3^{\mathrm{scr}}=10^{-3}~\mathrm{cm}^{-1}$, the candidate is added to $\mathcal{S}_{\mathrm{work}}$, and the enlarged-basis eigenvalues become the reference for the candidates considered thereafter. Once all candidates have been examined, another iteration is carried out if the current pass yielded any additions. Iterating in this manner captures states that become important only through indirect coupling to previously selected ones. The loop terminates when an iteration yields no further additions. The selected quantum-number list $\mathbf n(k)$ then defines the pruned VCI basis entering the final RVCI calculation.

\section{Results and Discussion}

In this section, we characterize the practical performance of the proposed basis pruning scheme for two systems. We first consider formaldehyde (\ch{H2CO}), for which polyad schemes have been employed extensively to truncate the vibrational basis in rovibrational calculations \cite{Yachmenev2011, Refaie015}, thereby offering an established alternative for comparison. Using a large-basis reference calculation, we assess the accuracy achieved with polyad truncation and our pruning in terms of rovibrational state energies and transition frequencies.

We then turn to thioketene (\ch{H2CCS}), for which no comparable polyad model is available. Here, we examine the convergence of the calculated energies and frequencies toward the large-basis reference results as the screening threshold is tightened. Since \ch{H2CCS} also represents a computationally more demanding system, it further allows us to quantify the associated savings in memory and computational time alongside the reduction in basis size.

\subsection{Formaldehyde (\ch{H2CO})}
\label{subsec:h2co}

All \ch{H2CO} calculations reported here are based on the same PES, VCI correlation space, and RVCI setup, as detailed in the Supplementary Material (SM), with only the composition of the vibrational basis being varied. The large-basis reference comprises 82 VCI states, including the vibrational ground state, with excitation levels up to four and energies below \(6000~\mathrm{cm}^{-1}\). Since the discussion below is organized by spectral regions, the six fundamentals are labeled \(1^1\)--\(6^1\) in order of increasing frequency rather than according to the conventional symmetry-based numbering, with \(i^n\) representing \(n\) quanta in mode \(i\). The correspondence between both labeling schemes, together with the VCI frequencies, is provided in the SM.

One common way to organize the vibrational state space of \ch{H2CO} is through a polyad number \(P\) \cite{Yachmenev2011,Refaie015,Rey2022NovelAnalysis,Nikitin2024,Rodina2026} (cf.\ Section \ref{sec:introduction}), defined as \cite{Refaie015,Rey2022NovelAnalysis}
\begin{equation}
    P=n_1+n_2+n_3+n_4+2(n_5+n_6),
\end{equation}
where \(n_i\) denotes the number of quanta in mode \(i\) (cf.\ Section~\ref{sec:methodology}). In line with the aforementioned convention, the modes are ordered by increasing frequency. Accordingly, the C--H stretching contributions \(n_5\) and \(n_6\) enter with twice the weight of \(n_1\)--\(n_4\), reflecting the approximately doubled fundamental frequencies of these modes. Choosing a maximum polyad number \(P_{\max}\) then provides a systematic criterion for selecting a reduced set of vibrational states for the subsequent rovibrational calculation \cite{Yachmenev2011,Refaie015}.

In particular, we consider three polyad-truncated bases, \(P_1\), \(P_2\), and \(P_3\), obtained by retaining all VCI states with \(P\leq1\), \(P\leq2\), and \(P\leq3\), respectively, as determined from their assigned vibrational quantum numbers \cite{Mathea2020}. These contain 5, 17, and 45 vibrational states and thus reduce the 82-state basis by 93.9\%, 79.3\%, and 45.1\%. Note that the comparatively low values of \(P_{\max}\) arise because the restriction is applied here to correlated VCI eigenstates rather than to primitive vibrational product functions \cite{Yachmenev2011,Refaie015}. Basis sets with \(P_{\max}\geq4\) would require additional vibrational states beyond those included in the reference, so that \(P_3\) represents the largest complete polyad still fully contained within this space.

Following Section~\ref{sec:methodology}, the pruning procedure is applied separately to three monitored spectral windows spanning \(500\)--\(1550~\mathrm{cm}^{-1}\), \(1550\)--\(2000~\mathrm{cm}^{-1}\), and \(2000\)--\(3000~\mathrm{cm}^{-1}\). These contain the target fundamentals \(1^1\)--\(3^1\), \(4^1\), and \(5^1\)--\(6^1\), respectively, with the remaining states of the 82-state reference basis serving as candidates. The three windows are screened using the thresholds \(Q_3^{\mathrm{pre}}=10^{-1}~\mathrm{cm}^{-1}\) and \(Q_3^{\mathrm{scr}}=10^{-3}~\mathrm{cm}^{-1}\). The resulting vibrational bases comprise 32, 23, and 35 VCI states, representing basis reductions of 61.0\%, 72.0\%, and 57.3\%, with all three sizes lying between those of \(P_2\) and \(P_3\). Table~\ref{tab:h2co_pruning_windows} summarizes the pruning sequence from the initial basis through preselection and the main pruning loop. Starting from only 4, 2, and 3 states, the main pruning loop converges after two iterations in every case.

\begin{table}[t]
\caption{\label{tab:h2co_pruning_windows}
Details of the three \ch{H2CO} pruning calculations (cf.\ Section~\ref{sec:methodology}). \(N_{\mathrm{init}}\) denotes the initial basis size, \(N_{\mathrm{pre}}\) and \(N_{\mathrm{scr}}\) the numbers of states added during preselection and screening, and \(N_{\mathrm{it}}\) the number of iterations to convergence. All counts include the vibrational ground state.}
\centering
\setlength{\tabcolsep}{4.5pt}
\renewcommand{\arraystretch}{1.00}
\begin{tabular}{cccccc}
\hline\hline
\shortstack{Window} &
\shortstack{Spectral range / \(\mathrm{cm}^{-1}\)} &
\(N_{\mathrm{init}}\) &
\(N_{\mathrm{pre}}\) &
\(N_{\mathrm{scr}}\) &
\(N_{\mathrm{it}}\) \\
\hline
1 & 500--1550  & 4 & 1 & 27 & 2 \\
2 & 1550--2000 & 2 & 2 & 19 & 2 \\
3 & 2000--3000 & 3 & 5 & 27 & 2 \\
\hline\hline
\end{tabular}
\end{table}

\begin{figure*}[t]
    \centering
    \includegraphics[width=\textwidth]{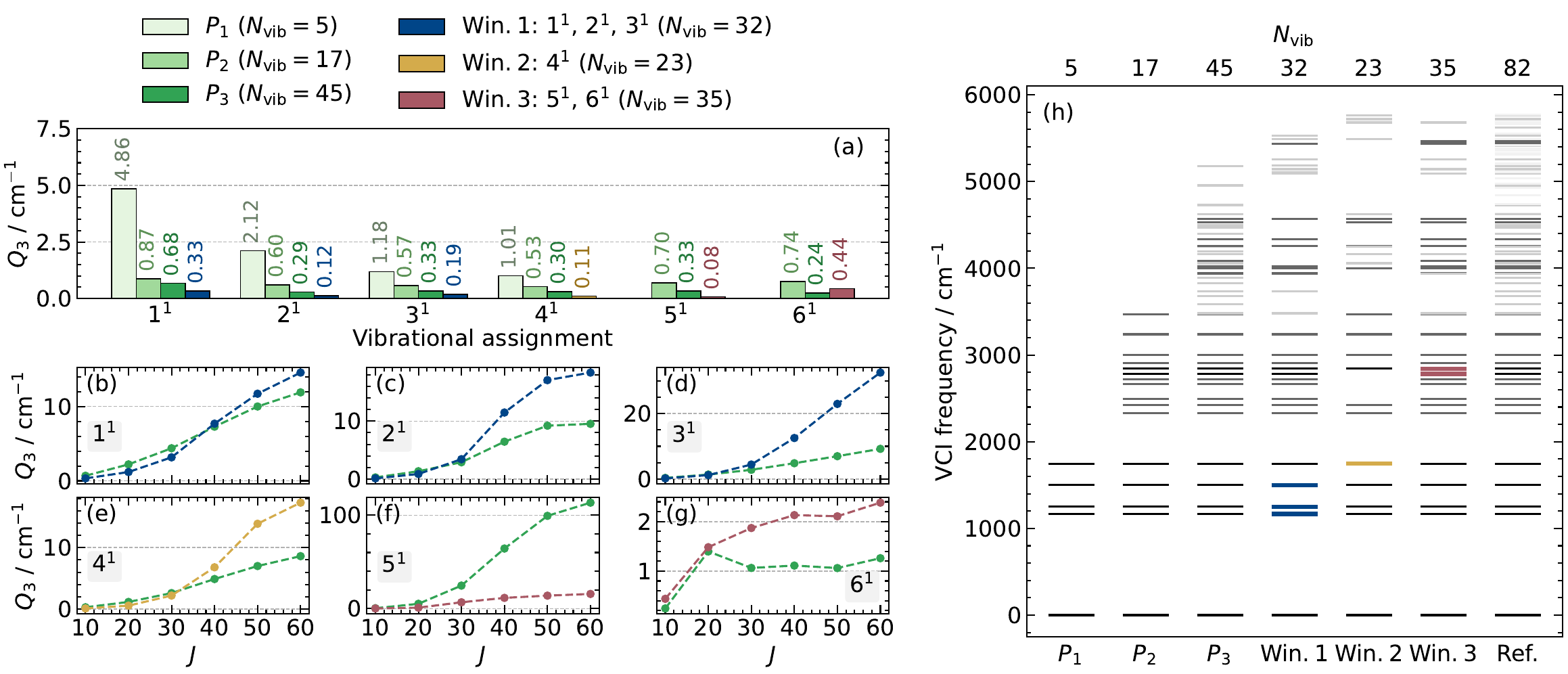}
    \caption{\label{fig:h2co_pruning_performance_en}
    Performance of the reduced vibrational bases for rovibrational states assigned to the six fundamentals of \ch{H2CO}. (a) Third quartile \(Q_3\) of the absolute energy deviations from the 82-state reference calculation for \(J\leq10\), comparing the polyad-truncated \(P_1\), \(P_2\), and \(P_3\) with the pruned bases obtained for windows 1 (\(500\)--\(1550~\mathrm{cm}^{-1}\), \(1^1\)--\(3^1\)), 2 (\(1550\)--\(2000~\mathrm{cm}^{-1}\), \(4^1\)), and 3 (\(2000\)--\(3000~\mathrm{cm}^{-1}\), \(5^1\)--\(6^1\)). (b)--(g) Same analysis as in (a), comparing \(P_3\) with the relevant pruned basis as progressively higher rotational quantum numbers are included up to \(J_{\max}=60\). At each point, \(Q_3\) is determined from all matched levels up to the indicated value of \(J\). (h) Composition of the vibrational bases. Horizontal lines mark the VCI energies of the retained states, with the gray scale becoming lighter with increasing total excitation order and the monitored fundamentals highlighted according to their pruning windows. The numbers above the columns give the corresponding basis sizes \(N_{\mathrm{vib}}\).}
\end{figure*}

To quantify how the different vibrational basis truncations affect the rovibrational state energies and transition frequencies associated with the six fundamentals of \ch{H2CO}, we compare the resulting values with the large-basis reference. Since a meaningful identification of equivalent levels depends on a reliable quantum-number assignment \cite{Tschoepe2022}, only states with an assignment confidence \(A\geq0.5\) are retained. Ambiguous duplicate assignments are removed, and the remaining levels are matched according to their quantum numbers. The resulting absolute energy deviations are summarized, in analogy to the pruning metric introduced in Section~\ref{subsec:pruning_metrics}, by the third quartile \(Q_3\), below which 75\% of the values lie. For the transition frequencies, only the six fundamental bands, comprising transitions from the vibrational ground state to the respective fundamentals, are considered, thereby excluding hot bands. After removing ambiguous duplicate assignments from the line lists \cite{Erfort2020}, the transitions are matched through the quantum numbers of their lower and upper states. Their absolute frequency deviations are characterized by the same \(Q_3\) measure.

As the main pruning loop explicitly monitors rovibrational eigenvalues only for \(J\leq10\), we first assess the state energies within this range. The analysis is then extended stepwise to \(J_{\max}=60\), with \(Q_3\) recalculated from all matched levels up to each value of \(J\), to examine how well the selected vibrational basis extrapolates beyond the screening range. The state-energy analysis is subsequently complemented by a comparison of transition frequencies for all matched transitions involving states with \(J\leq60\). Representative numbers of matched levels and transitions underlying these statistics are given throughout the discussion, while the complete statistics for the individual filtering and matching steps are compiled in the SM. Since the \ch{H2CO} reference calculation is already computationally inexpensive, computational savings are discussed instead for the more demanding \ch{H2CCS} case (cf.\ Section~\ref{subsec:h2ccs}).

\begin{figure*}[t]
    \centering
    \includegraphics[width=\textwidth]{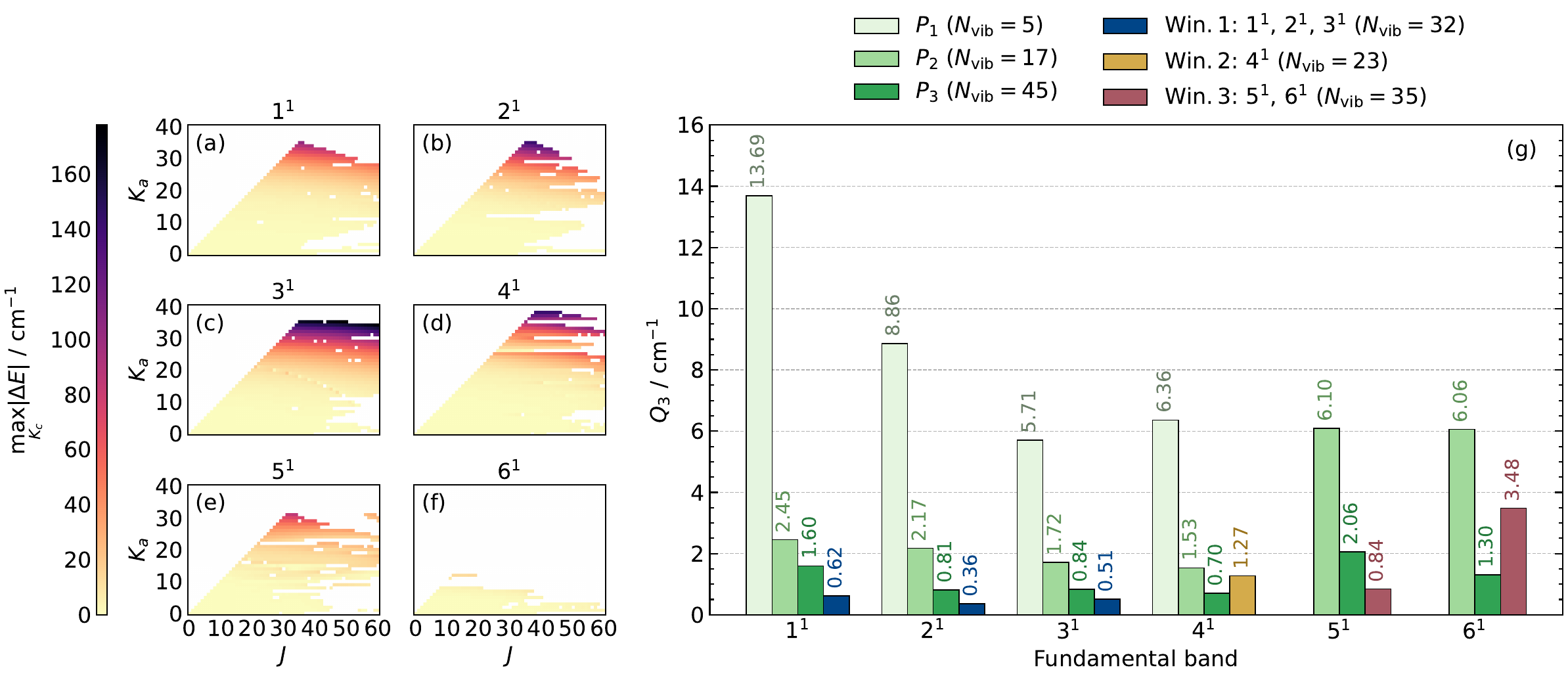}
    \caption{\label{fig:h2co_heatmaps_transition_q3}
    Rovibrational state-energy and transition-frequency deviations for the six fundamentals of \ch{H2CO}. (a)--(f) Absolute differences in the state energies between the pruned and 82-state reference calculations, mapped onto the \((J,K_a)\) plane. For \((J,K_a)\) pairs with two \(K_c\) components, the larger absolute deviation is shown. White regions indicate levels excluded by the \(A\geq0.5\) assignment-confidence criterion or lacking a unique match to the reference. (g) Third quartile \(Q_3\) of the absolute transition-frequency deviations for matched transitions involving states with \(J\leq60\), comparing the polyad-truncated \(P_1\), \(P_2\), and \(P_3\) with the pruned bases obtained for windows 1--3.}
\end{figure*}

Figure~\ref{fig:h2co_pruning_performance_en}(a) depicts the third quartile \(Q_3\) of the absolute rovibrational state-energy deviations for \(J\leq10\), grouped by vibrational assignment. For the polyad-truncated bases, the errors decrease systematically from \(P_1\) to \(P_3\) for all six fundamentals, as expected with increasing basis size. The particularly poor performance of \(P_1\), which contains only the vibrational ground state and the four lower-frequency fundamentals \(1^1\)--\(4^1\), shows that such a minimal vibrational basis is insufficient for converging RVCI state energies \cite{Tschoepe2022}. Despite containing fewer VCI states, the pruned bases improve on the \(P_3\) result by factors of about 2--4 for \(1^1\)--\(5^1\). The sole exception is the \(6^1\) fundamental, for which the pruned basis yields a larger deviation than \(P_3\).

The composition of the different vibrational bases is shown in Figure~\ref{fig:h2co_pruning_performance_en}(h). Whereas the polyad truncations follow a predefined hierarchy, the pruning procedure is designed to retain vibrational states according to their influence on the rovibrational energies within the monitored spectral window. Consequently, several overtone and combination states absent from \(P_3\) are selected, while other states included by the polyad criterion are omitted. Thus, we attribute the improved accuracy of the smaller pruned bases to this more selective construction.

The case of \(6^1\) merits a closer look. Although \(5^1\) and \(6^1\) belong to the same pruning window, their rovibrational state-energy deviations differ markedly, with \(Q_3=0.08~\mathrm{cm}^{-1}\) for \(5^1\) (\(P_3\): \(0.33~\mathrm{cm}^{-1}\)) and \(0.44~\mathrm{cm}^{-1}\) for \(6^1\) (\(P_3\): \(0.24~\mathrm{cm}^{-1}\)). An indication of the origin of this behavior is provided by the assignment statistics. For \(1^1\)--\(5^1\), all 121 expected levels pass the \(A\geq0.5\) confidence threshold for \(J\leq10\), whereas only 115 do so for \(6^1\). Since assignment confidence is known to decrease with increasing coupling strength \cite{Tschoepe2022}, this points to stronger mixing of the \(6^1\) levels with nearby states. In such a case, the influence on a given rovibrational level can be distributed over a larger set of coupled vibrational states. Some of these contributions may produce only small individual energy shifts and therefore remain below \(Q_3^{\mathrm{scr}}\) during the iterative selection, even though their combined effect can improve the final energies. The broader \(P_3\) space contains more of these potential interaction partners by construction, which may explain its lower \(Q_3\) value for \(6^1\). Importantly, this behavior can be addressed directly within the pruning scheme by tightening \(Q_3^{\mathrm{scr}}\), thereby retaining smaller contributions when higher accuracy is required.

Having established the performance within the \(J\leq10\) screening range, we next examine how the selected bases extrapolate to higher \(J\) in panels~(b)--(g) of Figure~\ref{fig:h2co_pruning_performance_en}. For \(1^1\)--\(4^1\), the pruned bases extrapolate well beyond the screening range at first, but their deviations grow more rapidly with increasing \(J\) than those for \(P_3\), reaching \(Q_3\) values of approximately \(15\)--\(33~\mathrm{cm}^{-1}\) for \(J\leq60\). A different trend is observed for \(5^1\), where pruning remains substantially closer to the reference, with \(Q_3=15.6~\mathrm{cm}^{-1}\) compared with \(113.4~\mathrm{cm}^{-1}\) for \(P_3\) at the \(J\leq60\) cutoff. For \(6^1\), both approaches remain at much smaller deviations and become nearly constant at higher \(J\), reaching \(2.38~\mathrm{cm}^{-1}\) and \(1.26~\mathrm{cm}^{-1}\) for pruning and \(P_3\), respectively (\textit{vide infra}). The numbers of matched levels underlying the individual statistics are provided in the SM.

These trends demonstrate that the importance of individual vibrational states changes with \(J\) through rotational and Coriolis couplings. Note that this behavior motivated the preselection discussed in Section~\ref{sec:methodology}. Crucially, the accuracy at higher \(J\) can be controlled through two parameters of the pruning sequence according to the requirements of the application. Tightening \(Q_3^{\mathrm{scr}}\) retains progressively smaller contributions to the monitored energies and thus improves convergence of the selected basis (cf.\ Section~\ref{subsec:h2ccs}). Alternatively, extending the main pruning loop to larger \(J\) allows interactions that become relevant there to enter the screening metric, albeit at increased computational cost. To further assess the origin and significance of the observed deviations at higher \(J\), we next analyze their distribution with respect to the rotational quantum numbers \(J\) and \(K_a\).

For this purpose, we map the absolute difference in state energies between the pruned and reference calculations onto the \((J,K_a)\) plane in Figure~\ref{fig:h2co_heatmaps_transition_q3}(a)--(f). When two \(K_c\) components are available for a given \((J,K_a)\) pair, the larger absolute deviation is shown. Since the corresponding values were found to differ only slightly, this provides a conservative representation without affecting the overall pattern. For the rovibrational states associated with \(1^1\)--\(5^1\), the largest deviations occur predominantly at high \(K_a\) and consequently among the higher-\(J\) levels. Notably, they are associated mainly with low assignment confidence. Such entries would typically be excluded, as is done, for instance, when RVCI state energies are filtered according to assignment quality before spectroscopic fitting \cite{Dinu2022}. A distinct pattern is observed for \(6^1\) in panel~(f), where the map becomes considerably more sparse because fewer levels satisfy the \(A\geq0.5\) confidence criterion. Specifically, up to \(J=60\), only 564 states remain in the matched set, compared with roughly \(2000\)--\(2700\) for the other fundamentals (cf.\ SM). The absence of precisely these high-\(K_a\) levels at larger \(J\) from the statistics also explains why the \(Q_3\) values for \(6^1\) remain comparatively small in Figure~\ref{fig:h2co_pruning_performance_en}(g).

We subsequently investigate how the deviations in rovibrational state energies carry over to the line positions in the spectrum. Panel~(g) of Figure~\ref{fig:h2co_heatmaps_transition_q3} presents the \(Q_3\) statistics for matched transitions involving states with \(J\leq60\), each based on approximately \(5000\)--\(7800\) transitions. Overall, the frequency deviations are substantially smaller than the energy errors over the same \(J\) range. As anticipated, the polyad-truncated bases improve progressively from \(P_1\) to \(P_3\) for all six fundamental bands. Pruning yields \(Q_3<1~\mathrm{cm}^{-1}\) for \(1^1\)--\(3^1\) and \(5^1\), improving on \(P_3\) by about a factor of two. The remaining two bands depart from this trend. The 23-state pruned basis gives \(Q_3=1.27~\mathrm{cm}^{-1}\) for \(4^1\), compared with \(0.70~\mathrm{cm}^{-1}\) from the 45-state \(P_3\) space. For \(6^1\), pruning and \(P_3\) yield \(3.48~\mathrm{cm}^{-1}\) and \(1.30~\mathrm{cm}^{-1}\), respectively, in line with the behavior identified above for this fundamental. In both cases, the pruned line lists also contain a larger fraction of duplicate assignments and yield fewer uniquely matched transitions than \(P_3\), with these differences being considerably more pronounced than in the four bands where pruning is more accurate (cf.\ SM). As noted above, higher accuracy can be targeted by tightening \(Q_3^{\mathrm{scr}}\), whose effect is examined systematically for \ch{H2CCS} in Section~\ref{subsec:h2ccs}.

The smaller errors in the line positions compared with the state energies for \(J\leq60\) can be rationalized by inspecting the rovibrational states associated with the largest energy deviations. Among the five largest energy shifts for each of the \(1^1\)--\(5^1\) fundamentals, none gives rise to a transition contained in the line list. Since only transitions with intensities exceeding \(10^{-6}\) relative to the strongest line are included, such weak transitions are omitted. Of the five most strongly shifted \(6^1\) levels, only two contribute to the line list, and even the stronger corresponding transition reaches merely \(0.24\%\) of the maximum intensity. Hence, the largest state-energy deviations predominantly concern levels that contribute only weakly, if at all, to the simulated spectra. Consistent with this finding, the stick spectra obtained with the pruned and reference bases are virtually indistinguishable on the scale of the complete bands and are therefore provided in the SM for completeness. More pronounced changes in the band shapes can, however, arise upon further reduction of the vibrational basis (cf.\ Section~\ref{subsec:h2ccs}).

Taken together, the \ch{H2CO} benchmark shows that the pruned bases reproduce the targeted state energies within the screening range and transition frequencies with an accuracy comparable to, and in most cases exceeding, that of larger polyad-truncated bases. This improvement reflects the selection of vibrational states according to their influence on the spectral region of interest, with the balance between basis size and accuracy being systematically adjustable through the screening threshold.

\subsection{Thioketene (\ch{H2CCS})}\label{subsec:h2ccs}
Having established the performance of the approach for \ch{H2CO}, we now turn to \ch{H2CCS}, for which, to the best of our knowledge, no suitable polyad scheme has been reported. Its higher computational cost also allows the computational savings achieved through basis pruning to be assessed more meaningfully. In analogy to the previous section, the PES, VCI correlation space, and RVCI setup are summarized in the SM. The fundamentals are again labeled in order of increasing frequency, with the conventional symmetry-based numbering and VCI frequencies listed in the SM.

The reference calculation used throughout this section employs a basis of 102 VCI states with energies below \(3500~\mathrm{cm}^{-1}\) and a maximum total excitation level of three. We choose a spectral range from \(150\) to \(1000~\mathrm{cm}^{-1}\) for pruning (cf.\ Section~\ref{sec:methodology}), which contains the five target fundamentals \(1^1\)--\(5^1\). Their associated rovibrational levels are monitored upon addition of candidate states drawn from the remainder of the 102-state reference basis. To characterize the convergence with respect to \(Q_3^{\mathrm{scr}}\), the screening threshold is varied from \(10^{-1}\) to \(10^{-5}~\mathrm{cm}^{-1}\), while \(Q_3^{\mathrm{pre}}\) is kept fixed at \(10^{-1}~\mathrm{cm}^{-1}\). Together with the vibrational ground state, the monitored fundamentals define the initial basis, to which preselection adds 14 states in every calculation. The main pruning loop therefore starts from a common 20-state working basis and selects additional states according to \(Q_3^{\mathrm{scr}}\). Table~\ref{tab:h2ccs_pruning} reports these additions together with the resulting basis size used in the subsequent RVCI calculation, its reduction relative to the reference, and the number of iterations required for convergence. The calculations at the two loosest thresholds converge after two iterations, whereas three are needed from \(Q_3^{\mathrm{scr}}=10^{-3}~\mathrm{cm}^{-1}\) onward.

\begin{table}[t]
    \caption{Details of the \ch{H2CCS} pruning calculations for different screening thresholds \(Q_3^{\mathrm{scr}}\). The initial basis contains six states, while preselection adds 14 in every calculation. \(N_{\mathrm{scr}}\) denotes the number of states added during the main pruning loop, \(N_{\mathrm{vib}}\) the resulting basis size, and \(N_{\mathrm{it}}\) the number of iterations required for convergence. The reduction is given relative to the 102-state reference. All reported basis sizes include the vibrational ground state.}
    \label{tab:h2ccs_pruning}
    \begin{ruledtabular}
    \begin{tabular}{ccccc}
        \(Q_3^{\mathrm{scr}}/\mathrm{cm}^{-1}\) &
        \(N_{\mathrm{scr}}\) &
        \(N_{\mathrm{vib}}\) &
        Reduction/\% &
        \(N_{\mathrm{it}}\) \\
        \hline
        \(10^{-1}\) & 1  & 21 & 79.4 & 2 \\
        \(10^{-2}\) & 10 & 30 & 70.6 & 2 \\
        \(10^{-3}\) & 24 & 44 & 56.9 & 3 \\
        \(10^{-4}\) & 43 & 63 & 38.2 & 3 \\
        \(10^{-5}\) & 64 & 84 & 17.6 & 3 \\
    \end{tabular}
    \end{ruledtabular}
\end{table}

\begin{figure*}[t]
    \centering
    \includegraphics[width=\textwidth]{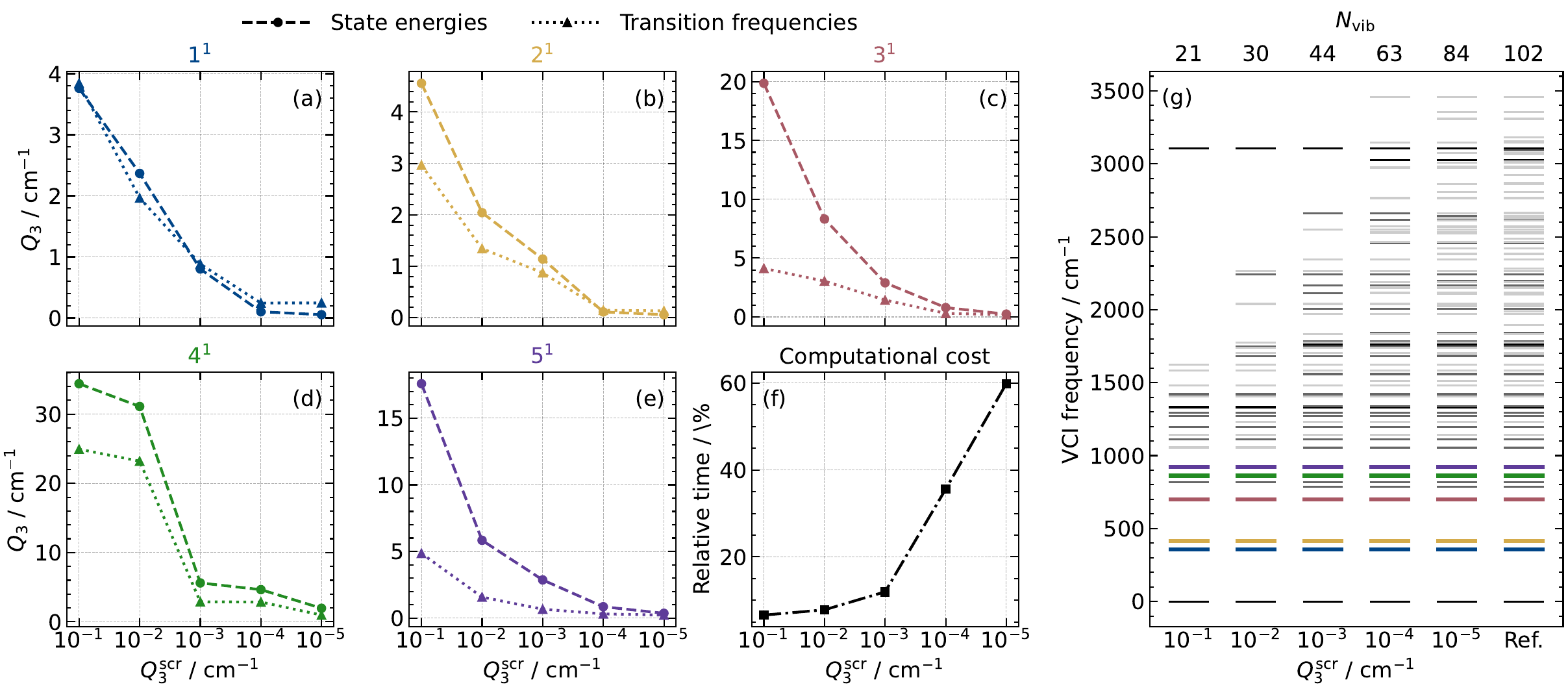}
    \caption{\label{fig:h2ccs_pruning_convergence} Convergence and computational performance of the pruned vibrational bases for \ch{H2CCS} as a function of the screening threshold \(Q_3^{\mathrm{scr}}\). (a)--(e) Third quartile \(Q_3\) of the absolute rovibrational state-energy and transition-frequency deviations from the 102-state reference for the five target fundamentals \(1^1\)--\(5^1\) at \(J\leq110\). (f) Relative total computation time, obtained by normalizing \(t_{\rm calc}\) to the reference calculation. Here, \(t_{\rm calc}\) represents the combined time required for the pruning sequence and subsequent VCI and RVCI calculations up to \(J_{\max}=110\). The underlying absolute calculation times are given in Table~\ref{tab:h2ccs_timings}. (g) Composition of the vibrational bases. Horizontal lines indicate the VCI energies, with the gray scale becoming lighter with increasing total excitation order and the target fundamentals highlighted in color. The numbers above the columns indicate the basis sizes \(N_{\mathrm{vib}}\). } 
\end{figure*}

Figure~\ref{fig:h2ccs_pruning_convergence}(a)--(e) shows the convergence of the rovibrational state energies and transition frequencies for the five monitored fundamentals at \(J\leq110\). As in Section~\ref{subsec:h2co}, only reliably assigned and uniquely matched levels and transitions are included, and their absolute deviations from the 102-state reference are summarized by \(Q_3\). The numbers entering the respective comparisons are provided in the SM. Tightening \(Q_3^{\mathrm{scr}}\) consistently reduces the deviations across all five fundamentals. At \(10^{-1}~\mathrm{cm}^{-1}\), the state-energy \(Q_3\) values amount to only \(3.76\) and \(4.56~\mathrm{cm}^{-1}\) for \(1^1\) and \(2^1\), increase to about \(18\)--\(20~\mathrm{cm}^{-1}\) for \(3^1\) and \(5^1\), and reach \(34.40~\mathrm{cm}^{-1}\) for \(4^1\). The transition frequencies are less sensitive for four of the five fundamentals, with \(Q_3\) values between \(2.97\) and \(4.85~\mathrm{cm}^{-1}\), while \(4^1\) again stands out at \(24.93~\mathrm{cm}^{-1}\). At \(Q_3^{\mathrm{scr}}=10^{-3}~\mathrm{cm}^{-1}\), all state-energy and transition-frequency deviations have fallen below \(5.61\) and \(2.88~\mathrm{cm}^{-1}\), respectively. \\ \indent The modest state-energy deviations in Figure~\ref{fig:h2ccs_pruning_convergence}(a)--(e), despite extending the analysis to \(J\leq110\), can be understood from the assignment filtering. For \ch{H2CO}, the largest errors were found predominantly for high-\(K_a\) levels with \(K_a\) approaching \(J\), where increasingly ambiguous assignments limit the \(Q_3\) statistics, most notably for \(6^1\). This effect is considerably more widespread in \ch{H2CCS}. Depending on the fundamental, only about \(16\)--\(32\%\) of the levels of the 102-state reference satisfy \(A\geq0.5\) and can be uniquely matched, compared with approximately \(57\)--\(78\%\) for \(1^1\)--\(5^1\) of \ch{H2CO} over \(J\leq60\). We note that, although assignment confidence decreases with increasing state mixing, such filtering is necessary because a meaningful state-by-state comparison requires an unambiguous mapping of the rovibrational levels, as also discussed in Ref.~\onlinecite{Tschoepe2022}. In analogy to Figure~\ref{fig:h2co_heatmaps_transition_q3}(a)--(f), heatmaps for all screening thresholds are included in the SM, where their pronounced sparsity reflects the reduced number of uniquely identifiable levels and the remaining entries illustrate how the energy errors approach the reference. \\ \indent For the transition frequencies, the \(Q_3\) values are typically smaller than those of the state energies, following the behavior identified for \ch{H2CO}. Inspection of the five largest matched state-energy errors for every fundamental and screening threshold confirms the same origin, as only very few of these levels contribute to the line lists and the associated transitions reach at most about \(0.02\%\) of the maximum band intensity in the cases examined. Where such a level does contribute, its frequency error closely follows the underlying energy shift, consistent with the similar numerical values observed for several points in Figure~\ref{fig:h2ccs_pruning_convergence}(a)--(e). The corresponding stick spectra for the monitored low-frequency region are provided in the SM. We return to the pronounced threshold dependence of the \(4^1\) fundamental below. \\ \indent Panel~(g) illustrates how the composition of the vibrational basis evolves as \(Q_3^{\mathrm{scr}}\) is tightened. The basis grows from 21 to 84 states, with states entering over a broad range of VCI energies rather than according to energetic ordering alone, consistent with the behavior observed in Figure~\ref{fig:h2co_pruning_performance_en}(h). This increase in basis size is accompanied by a substantial rise in computational cost. Table~\ref{tab:h2ccs_timings} summarizes the time required for the complete pruning sequence, \(t_{\mathrm{prune}}\), and the total calculation time \(t_{\mathrm{calc}}\), which includes both the pruning procedure and the subsequent VCI and RVCI calculations. As discussed in Section~\ref{sec:theory}, increasing \(N_{\mathrm{vib}}\) enlarges the underlying VCI and RVCI eigenvalue problems, while the higher memory requirements of the largest calculations can additionally restrict the degree of parallelization. \\ \indent Upon tightening \(Q_3^{\mathrm{scr}}\), \(t_{\mathrm{prune}}\) increases from 49~min to 2~h~31~min, while \(t_{\mathrm{calc}}\) grows more strongly from 1~h~43~min to 15~h~37~min. The pruning step thus constitutes a decreasing fraction of the overall computational cost as the basis is enlarged. The relative computational cost is shown in Figure~\ref{fig:h2ccs_pruning_convergence}(f), which presents \(t_{\mathrm{calc}}\) relative to the 26~h~05~min required by the 102-state reference calculation. The relative time increases from 6.6\% at \(Q_3^{\mathrm{scr}}=10^{-1}~\mathrm{cm}^{-1}\) to 59.9\% at \(10^{-5}~\mathrm{cm}^{-1}\), while the default threshold of \(10^{-3}~\mathrm{cm}^{-1}\) requires only 12.0\%, corresponding to 3~h~07~min and a more than eightfold reduction. Together with the convergence observed in panels~(a)--(e), this provides a favorable compromise between accuracy and computational cost for the present \ch{H2CCS} calculations.

\begin{table}[t]
\caption{Computational effort of the \ch{H2CCS} calculations for the different screening thresholds \(Q_3^{\mathrm{scr}}\) and respective vibrational basis sizes \(N_{\mathrm{vib}}\). Timings are shown for the pruning step, \(t_{\mathrm{prune}}\), and for the total calculation, \(t_{\mathrm{calc}}\), which combines the pruning time with the subsequent VCI and RVCI calculations up to \(J_{\max}=110\).}
\label{tab:h2ccs_timings}
\begin{ruledtabular}
\begin{tabular}{cccc}
\(Q_3^{\mathrm{scr}}/\mathrm{cm}^{-1}\) &
\(N_{\mathrm{vib}}\) &
\(t_{\mathrm{prune}}\) &
\(t_{\mathrm{calc}}\) \\
\hline
\(10^{-1}\) & 21  & 49 min      & 1 h 43 min  \\
\(10^{-2}\) & 30  & 52 min      & 2 h 02 min  \\
\(10^{-3}\) & 44  & 1 h 21 min  & 3 h 07 min  \\
\(10^{-4}\) & 63  & 2 h 05 min  & 9 h 17 min  \\
\(10^{-5}\) & 84  & 2 h 31 min  & 15 h 37 min \\
Ref.        & 102 & --          & 26 h 05 min \\
\end{tabular}
\end{ruledtabular}
\end{table}

\begin{figure}[t]
\centering
\includegraphics[width=\columnwidth]{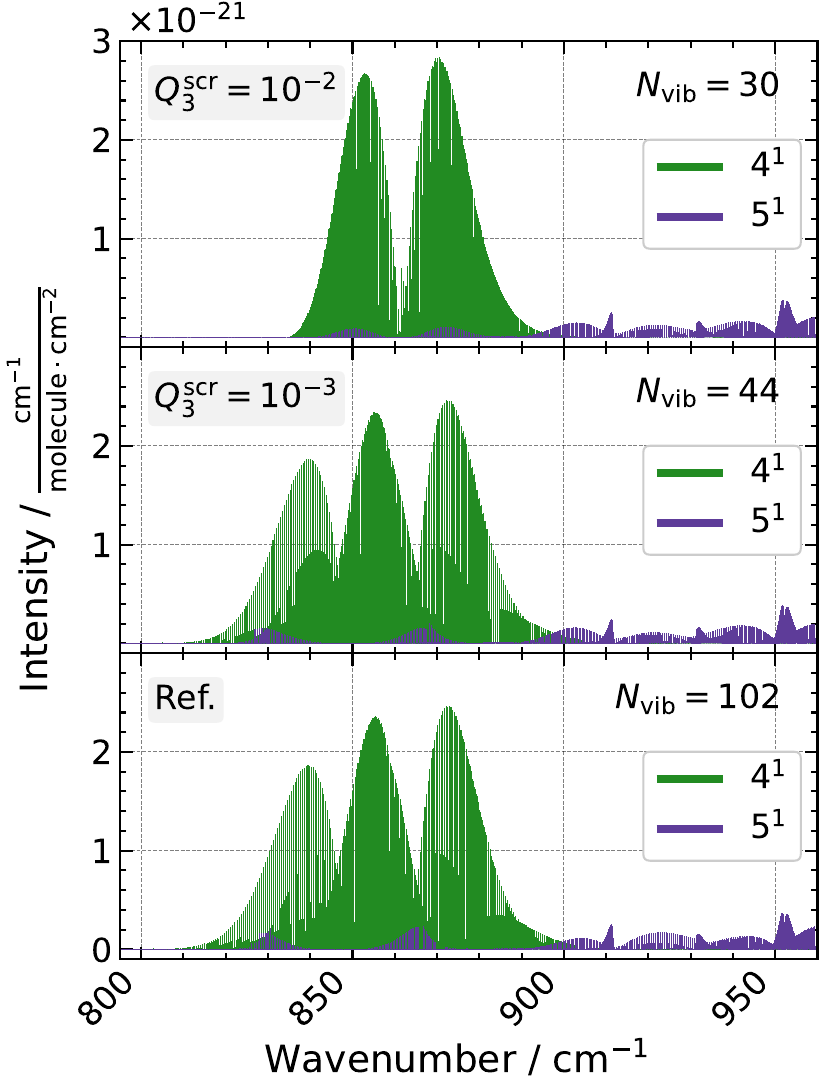}
\caption{\label{fig:h2ccs_41_51_spectra} Calculated \ch{H2CCS} RVCI stick spectra at \(T=300~\mathrm{K}\) between \(800\) and \(960~\mathrm{cm}^{-1}\), displaying only transitions whose upper states are assigned to the \(4^1\) and \(5^1\) fundamentals (cf.\ SM). Results are presented for \(Q_3^{\mathrm{scr}}=10^{-2}~\mathrm{cm}^{-1}\) (\(N_{\mathrm{vib}}=30\)), the default threshold \(Q_3^{\mathrm{scr}}=10^{-3}~\mathrm{cm}^{-1}\) (\(N_{\mathrm{vib}}=44\)), and the large-basis reference (\(N_{\mathrm{vib}}=102\)).}
\end{figure}

The rovibrational transitions associated with the \(4^1\) fundamental exhibit, as noted earlier, the largest deviations from the large-basis reference at the looser screening thresholds in panel~(d) of Figure~\ref{fig:h2ccs_pruning_convergence}. In particular, these deviations show a pronounced decrease between \(Q_3^{\mathrm{scr}}=10^{-2}\) and \(10^{-3}~\mathrm{cm}^{-1}\). This change is also apparent in the RVCI stick spectra at \(T=300~\mathrm{K}\) presented in Figure~\ref{fig:h2ccs_41_51_spectra}, where the results for these two thresholds are compared with the reference. At \(Q_3^{\mathrm{scr}}=10^{-2}~\mathrm{cm}^{-1}\), the \(4^1\) band exhibits a distinctly different structure, whereas the default value of \(10^{-3}~\mathrm{cm}^{-1}\) already reproduces the reference closely. Among the states admitted upon this tightening, the \(A_2\)-symmetric combination state \(1^12^1\) has a particularly strong influence on the band. Indeed, a reduced RVCI calculation containing only \(4^1\) and \(1^12^1\) produces the additional structure observed in the larger basis. In the first screening iteration, \(1^12^1\) yields \(Q_3=2.98\times10^{-3}~\mathrm{cm}^{-1}\) and is excluded at \(10^{-2}\) but retained at the default threshold. The known \(A_1\)-symmetric Fermi partner \(2^2\) of \(4^1\), on the other hand, is already selected during preselection. This case thus illustrates how tightening the screening threshold can recover an interaction partner whose inclusion has a pronounced effect on the resulting band shape.

\begin{figure}[t]
\centering
\includegraphics[width=\columnwidth]{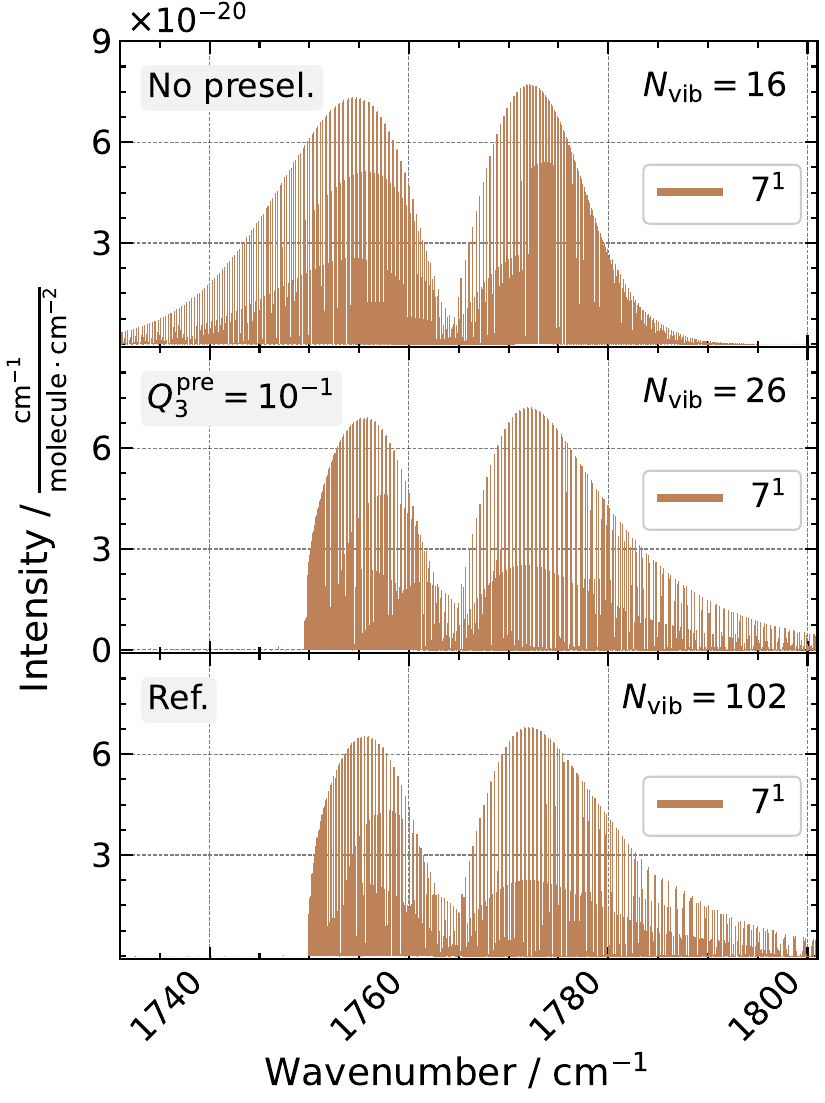}
\caption{\label{fig:h2ccs_preselection} Calculated \ch{H2CCS} RVCI stick spectra at \(T=300~\mathrm{K}\) between \(1730\) and \(1800~\mathrm{cm}^{-1}\), displaying only transitions whose upper states are assigned to the \(7^1\) fundamental (cf.\ SM). Results are presented without preselection (\(N_{\mathrm{vib}}=16\)), with preselection (\(N_{\mathrm{vib}}=26\)) at \(Q_3^{\mathrm{scr}}=10^{-2}~\mathrm{cm}^{-1}\), and for the large-basis reference (\(N_{\mathrm{vib}}=102\)).}
\end{figure}

The weaker transitions whose upper states are assigned to \(5^1\) warrant a separate comment. Below approximately \(900~\mathrm{cm}^{-1}\), where they overlap with the intense \(4^1\) band, several lines appear to shift as the vibrational basis is changed. In this strongly mixed region, however, the dominant effect is a redistribution of vibrational character among neighbouring eigenstates, which can alter their assignment to \(5^1\). Another aspect is visible in the \(5^1\) progression near \(920~\mathrm{cm}^{-1}\), where the intensity profile is sensitive to the vibrational basis. Such differences are not directly controlled by the pruning metric, which is defined exclusively from rovibrational energy shifts. Accounting for intensities during the iterative screening would require evaluating them for every tested basis \cite{Erfort2020} and would substantially increase the computational effort. Consequently, intensity convergence is best regarded as a separate aspect of the calculation. \\ \indent We conclude this section by considering the role of the preselection step for the \(7^1\) band. As motivated in Section~\ref{sec:methodology}, interactions that become important only at higher \(J\) may remain undetected during the low-\(J\) screening, yet still have a pronounced influence on the resulting band shape. For this purpose, the spectral window from \(1500\) to \(2000~\mathrm{cm}^{-1}\) was pruned using the relaxed threshold \(Q_3^{\mathrm{scr}}=10^{-2}~\mathrm{cm}^{-1}\), with further computational details given in the SM. Figure~\ref{fig:h2ccs_preselection} compares the resulting RVCI stick spectra at \(T=300~\mathrm{K}\). Without preselection, the 16-state basis fails to reproduce the characteristic band head of the large-basis reference. Including the preselected states increases the basis to only 26 states and already recovers the overall band shape. \\ \indent The effect originates from the nearby combination state \(2^24^1\), located about \(33~\mathrm{cm}^{-1}\) below \(7^1\). Both states have \(A_1\) symmetry, and their rotational interaction gives rise to an avoided crossing responsible for the band head. At \(Q_3^{\mathrm{scr}}=10^{-2}~\mathrm{cm}^{-1}\), the influence of \(2^24^1\) is too weak in the main pruning loop for the state to be retained, whereas the preselection calculation yields \(Q_3=3.37~\mathrm{cm}^{-1}\), well above \(Q_3^{\mathrm{pre}}=10^{-1}~\mathrm{cm}^{-1}\). Note that at the default screening threshold of \(Q_3^{\mathrm{scr}}=10^{-3}~\mathrm{cm}^{-1}\), the same interaction partner is also recovered by the iterative screening. The looser threshold used here therefore serves as a deliberate test of the preselection for interactions whose influence becomes significant at higher \(J\). \\ \indent In summary, the rovibrational state energies and transition frequencies converge toward the 102-state reference as \(Q_3^{\mathrm{scr}}\) is tightened. The associated increase in vibrational basis size raises the computational cost, allowing the screening threshold to control both the level of convergence and the required effort. The calculated spectra further demonstrate the importance of the basis composition, as omission of individual interaction partners can lead to pronounced changes in the predicted band structure. This is addressed by the preselection and iterative screening stages of the pruning scheme, which provide a systematic means of identifying such states according to their influence on the spectral region of interest. 

\section{Summary and Conclusions}
The question of how to choose an appropriate vibrational basis is a well-known problem in variational rovibrational calculations and is complicated by its dependence on both the molecule and the spectral region under consideration. In the context of rovibrational configuration interaction (RVCI), a basis-pruning scheme has been presented for the contracted vibrational basis generated by configuration-selective vibrational configuration interaction (cs-VCI). It relies on screening candidate vibrational states according to their influence on the rovibrational levels of interest. As computing a VCI wavefunction for every candidate would largely offset the computational benefit of pruning, this screening instead employs second-order vibrational M{\o}ller--Plesset (VMP2)-like wavefunctions constructed in reduced correlation spaces. To limit the cost of repeated RVCI diagonalizations, the main pruning loop is restricted to low rotational quantum numbers and enlarges the working basis iteratively whenever a candidate exceeds a given screening threshold. A preceding preselection targets selected state pairs whose interactions may become important only at higher rotational quantum numbers and probes them in one- and two-state RVCI calculations.

The formaldehyde benchmark showed that the pruned bases can yield smaller deviations from the large-basis reference in both state energies and transition frequencies than the polyad-truncated basis, despite containing fewer vibrational states. This comparison thus further illustrates that the suitability of a vibrational basis depends not only on its size, but also on selecting the states most relevant to the targeted rovibrational problem. For the states and transitions associated with the five lowest-frequency fundamentals of thioketene, tightening the screening threshold led to increasing agreement with the reference results. Moreover, even after accounting for the cost of the pruning procedure itself, the overall computation time for the pruning, VCI, and RVCI steps was reduced by nearly an order of magnitude relative to the large-basis reference. As expected, this computational advantage decreases as increasingly tight screening thresholds are required for higher accuracy.

The approach presented here constitutes one realization of a more general philosophy in which basis functions are selected based on their relevance to a defined target. Methodological choices such as the representation of candidate states or the screening metric can be adapted to a particular application, provided that the resulting gain remains commensurate with the additional computational effort. This flexibility makes the proposed pruning scheme an attractive alternative to established basis-truncation strategies and may facilitate its extension to other variational methods in theoretical rovibrational spectroscopy.

\section*{Supplementary Material}
The supplementary material contains additional computational details, assignment and matching statistics, and calculated spectra for formaldehyde and thioketene. Furthermore, the full set of state-energy deviation maps for thioketene is provided.

\section*{Author Declarations}
\subsection*{Conflict of Interest}
The authors have no conflicts to disclose.
\subsection*{Author Contributions}
\noindent
\textbf{K. Oenen}:
Conceptualization (supporting), Methodology (equal), Software (lead), Validation (lead), Formal Analysis (lead), Investigation (lead), Data Curation (lead), Writing – Original Draft Preparation (lead), Visualization (lead)\\
\textbf{S. Das}:
Methodology (supporting), Software (supporting), Investigation (supporting)\\
\textbf{G. Rauhut:}
Conceptualization (lead), Methodology (equal), Investigation (supporting), Resources (lead), Writing – Review \& Editing (lead), Supervision (lead), Project Administration (lead), Funding Acquisition (lead)
\section*{Data Availability}
The data that support the findings of this study, i.e., the potential energy surfaces and line lists of formaldehyde and thioketene, are available online in the {\sc Zenodo} database (https://www.doi.org/10.5281/zenodo.21792313).

\section*{Acknowledgments}
We thank Dr.~Dennis F.~Dinu for proofreading the manuscript and for helpful suggestions. Financial support by the Deutsche Forschungsgemeinschaft (Ra 656/32-1) is kindly acknowledged.
\bibliography{literature}

@article{Polyad_Review,
  author = {Herman, Michel and Perry, David S.},
  title = {Molecular spectroscopy and dynamics: a polyad-based perspective},
  journal = {Phys. Chem. Chem. Phys.},
  year = {2013},
  volume = {15},
  number = {25},
  pages = {9970--9993},
  publisher = {The Royal Society of Chemistry},
  doi = {10.1039/C3CP50463H}
}

@article{Schroeder2024,
  author = {Schr{\"o}der, Benjamin and Rauhut, Guntram},
  title = {From the {Automated} {Calculation} of {Potential} {Energy} {Surfaces} to {Accurate} {Infrared} {Spectra}},
  journal = {J. Phys. Chem. Lett.},
  volume = {15},
  number = {11},
  pages = {3159--3169},
  year = {2024},
  doi = {10.1021/acs.jpclett.4c00186}
}

@article{Das2024,
  author = {Das, Subhasish and Rauhut, Guntram},
  title = {Acceleration of rovibrational spectrum calculations through sparsity techniques},
  journal = {J. Chem. Phys.},
  volume = {161},
  number = {20},
  pages = {204101},
  year = {2024},
  doi = {10.1063/5.0240689}
}

@article{Mathea2020,
  author = {Mathea, Tina and Rauhut, Guntram},
  title = {Assignment of vibrational states within configuration interaction calculations},
  journal = {J. Chem. Phys.},
  volume = {152},
  number = {19},
  pages = {194112},
  year = {2020},
  doi = {10.1063/5.0009732}
}

@incollection{Schroeder2022,
  author = {Schr{\"o}der, Benjamin and Rauhut, Guntram},
  title = {Vibrational Configuration Interaction Theory},
  booktitle = {Vibrational Dynamics of Molecules},
  editor = {Bowman, Joel M.},
  chapter = {1},
  year = {2022},
  publisher = {World Scientific},
  pages = {1--40},
  doi = {10.1142/9789811237911_0001}
}

@article{Watson1968,
  author = {Watson, James K. G.},
  title = {Simplification of the molecular vibration-rotation {Hamiltonian}},
  journal = {Mol. Phys.},
  volume = {15},
  number = {5},
  pages = {479--490},
  year = {1968},
  doi = {10.1080/00268976800101381}
}

@article{Werner2020,
  author = {Werner, Hans-Joachim and Knowles, Peter J. and Manby, Frederick R. and Black, Joshua A. and Doll, Klaus and He{\ss}elmann, Andreas and Kats, Daniel and K{\"o}hn, Andreas and Korona, Tatiana and Kreplin, David A. and Ma, Qianli and Miller, Thomas F. and Mitrushchenkov, Alexander and Peterson, Kirk A. and Polyak, Iakov and Rauhut, Guntram and Sibaev, Marat},
  title = {The {Molpro} quantum chemistry package},
  journal = {J. Chem. Phys.},
  volume = {152},
  number = {14},
  pages = {144107},
  year = {2020},
  doi = {10.1063/5.0005081}
}

@article{Tschoepe2022,
  author = {Tsch{\"o}pe, Martin and Rauhut, Guntram},
  title = {Convergence of series expansions in rovibrational configuration interaction ({RVCI}) calculations},
  journal = {J. Chem. Phys.},
  volume = {157},
  pages = {234105},
  year = {2022},
  doi = {10.1063/5.0129828}
}

@article{Erfort2020,
  author = {Erfort, Sebastian and Tsch{\"o}pe, Martin and Rauhut, Guntram},
  title = {Toward a fully automated calculation of rovibrational infrared intensities for semi-rigid polyatomic molecules},
  journal = {J. Chem. Phys.},
  volume = {152},
  pages = {244104},
  year = {2020},
  doi = {10.1063/5.0011832}
}

@article{Mathea2021,
  author = {Mathea, Tina and Rauhut, Guntram},
  title = {Advances in vibrational configuration interaction theory -- part 1: {Efficient} calculation of vibrational angular momentum terms},
  journal = {J. Comput. Chem.},
  volume = {42},
  pages = {2321--2333},
  year = {2021},
  doi = {10.1002/jcc.26762}
}

@article{Mathea2022,
  author = {Mathea, Tina and Petrenko, Taras and Rauhut, Guntram},
  title = {Advances in vibrational configuration interaction theory -- part 2: {Fast} screening of the correlation space},
  journal = {J. Comput. Chem.},
  volume = {43},
  pages = {6--18},
  year = {2022},
  doi = {10.1002/jcc.26764}
}

@article{Ziegler2016,
  author = {Ziegler, B. and Rauhut, G.},
  title = {Efficient generation of sum-of-products representations of high-dimensional potential energy surfaces based on multimode expansions},
  journal = {J. Chem. Phys.},
  volume = {144},
  pages = {114114},
  year = {2016},
  doi = {10.1063/1.4943985}
}

@article{Hamilton1986,
  author = {Hamilton, I. P. and Light, J. C.},
  title = {On distributed {Gaussian} bases for simple model multidimensional vibrational problems},
  journal = {J. Chem. Phys.},
  volume = {84},
  pages = {306--317},
  year = {1986},
  doi = {10.1063/1.450139}
}

@article{Bowman1986,
  author = {Bowman, J. M.},
  title = {The self-consistent-field approach to polyatomic vibrations},
  journal = {Acc. Chem. Res.},
  volume = {19},
  pages = {202--208},
  year = {1986},
  doi = {10.1021/ar00127a002}
}

@article{Carter1997,
  author = {Carter, Stuart and Culik, S. J. and Bowman, Joel M.},
  title = {Vibrational self-consistent field method for many-mode systems: A new approach and application to the vibrations of {CO} adsorbed on {Cu}(100)},
  journal = {J. Chem. Phys.},
  volume = {107},
  pages = {10458--10469},
  year = {1997},
  doi = {10.1063/1.474210}
}

@article{Rauhut2007,
  author = {Rauhut, Guntram},
  title = {Configuration selection as a route towards efficient vibrational configuration interaction calculations},
  journal = {J. Chem. Phys.},
  volume = {127},
  pages = {184109},
  year = {2007},
  doi = {10.1063/1.2790016}
}

@article{Rauhut2004,
  author = {Rauhut, Guntram},
  title = {Efficient calculation of potential energy surfaces for the generation of vibrational wave functions},
  journal = {J. Chem. Phys.},
  volume = {121},
  pages = {9313--9322},
  year = {2004},
  doi = {10.1063/1.1804174}
}

@article{Neff2009,
  author = {Neff, M. and Rauhut, G.},
  title = {Toward large scale vibrational configuration interaction calculations},
  journal = {J. Chem. Phys.},
  volume = {131},
  pages = {124129},
  year = {2009},
  doi = {10.1063/1.3243862}
}

@article{Ziegler2018,
  author = {Ziegler, B. and Rauhut, G.},
  title = {Rigorous use of symmetry within the construction of multidimensional potential energy surfaces},
  journal = {J. Chem. Phys.},
  volume = {149},
  pages = {164110},
  year = {2018},
  doi = {10.1063/1.5047912}
}

@article{Neff2011,
  author = {Neff, M. and Hrenar, T. and Oschetzki, D. and Rauhut, G.},
  title = {Convergence of vibrational angular momentum terms within the {Watson} {Hamiltonian}},
  journal = {J. Chem. Phys.},
  volume = {134},
  pages = {064105},
  year = {2011},
  doi = {10.1063/1.3551513}
}

@article{Spirko1985,
  author = {{\v{S}}pirko, V. and Jensen, P. and Bunker, P. and {\v{C}}ejchan, A.},
  title = {The development of a new {Morse}-oscillator based rotation-vibration {Hamiltonian} for {H$_3^+$}},
  journal = {J. Mol. Spectrosc.},
  volume = {112},
  pages = {183--202},
  year = {1985},
  doi = {10.1016/0022-2852(85)90202-6}
}

@article{Yurchenko2005,
  author = {Yurchenko, S. N. and Carvajal, M. and Jensen, P. and Lin, H. and Zheng, J. and Thiel, W.},
  title = {Rotation-vibration motion of pyramidal {XY$_3$} molecules described in the {Eckart} frame: {Theory} and application to {NH$_3$}},
  journal = {Mol. Phys.},
  volume = {103},
  pages = {359--378},
  year = {2005},
  doi = {10.1080/00268970512331317336}
}

@article{Tschoepe2021,
  author = {Tsch{\"o}pe, Martin and Schr{\"o}der, Benjamin and Erfort, Sebastian and Rauhut, Guntram},
  title = {High-Level Rovibrational Calculations on Ketenimine},
  journal = {Front. Chem.},
  volume = {8},
  pages = {623641},
  year = {2021},
  doi = {10.3389/fchem.2020.623641}
}

@article{Schneider2022,
  author = {Schneider, Moritz and Rauhut, Guntram},
  title = {Quantum chemical rovibrational analysis of aminoborane and its isotopologues},
  journal = {J. Comput. Chem.},
  volume = {43},
  pages = {1370--1382},
  year = {2022},
  doi = {10.1002/jcc.26893}
}

@article{Erfort2022,
  author = {Erfort, Sebastian and Tsch{\"o}pe, Martin and Rauhut, Guntram},
  title = {Efficient and automated quantum chemical calculation of rovibrational nonresonant {Raman} spectra},
  journal = {J. Chem. Phys.},
  volume = {156},
  pages = {124102},
  year = {2022},
  doi = {10.1063/5.0086749}
}

@article{Tschoepe2023a,
  author = {Tsch{\"o}pe, Martin and Rauhut, Guntram},
  title = {A theoretical study of propynal under interstellar conditions and beyond, covering low-frequency infrared spectra, spectroscopic constants, and hot bands},
  journal = {Mon. Not. R. Astron. Soc.},
  volume = {520},
  pages = {3345--3354},
  year = {2023},
  doi = {10.1093/mnras/stad251}
}

@article{Tschoepe2023b,
  author = {Tsch{\"o}pe, Martin and Rauhut, Guntram},
  title = {Spectroscopic Characterization of Diazophosphane—A Candidate for Astrophysical Observations},
  journal = {Astrophys. J.},
  volume = {949},
  pages = {1},
  year = {2023},
  doi = {10.3847/1538-4357/acc9ad}
}

@article{DeVos2023,
  author = {De Vos, John and Schr{\"o}der, Benjamin and Rauhut, Guntram},
  title = {Comprehensive quantum chemical analysis of the (ro)vibrational spectrum of thiirane and its deuterated isotopologue},
  journal = {Spectrochim. Acta A},
  volume = {302},
  pages = {123083},
  year = {2023},
  doi = {10.1016/j.saa.2023.123083}
}

@article{Das2024b,
  author = {Das, Subhasish and Rauhut, Guntram},
  title = {Rovibrational calculations without model {Hamiltonians}: The infrared and microwave spectra of thiopropynal},
  journal = {Int. J. Quantum Chem.},
  volume = {124},
  pages = {e27378},
  year = {2024},
  doi = {10.1002/qua.27378}
}

@article{Das2024c,
  author = {Das, Subhasish and Tsch{\"o}pe, Martin and Rauhut, Guntram},
  title = {Thioformyl cyanide, {HC(S)CN}, revisited: accurate rovibrational simulations for a molecule observed in interstellar clouds},
  journal = {Mol. Phys.},
  volume = {122},
  pages = {e2262059},
  year = {2024},
  doi = {10.1080/00268976.2023.2262059}
}

@article{Das2024d,
  author = {Das, Subhasish and Rauhut, Guntram},
  title = {Hunting for Sulfur-containing Molecules in Space: A Spectroscopic Characterization of Cyclopropenethione Based on High-level Ab Initio Calculations},
  journal = {Astrophys. J.},
  volume = {974},
  pages = {73},
  year = {2024},
  doi = {10.3847/1538-4357/ad73a0}
}

@article{Oenen2025,
  author = {Oenen, Kemal and Das, Subhasish and Liedl, Klaus R. and Rauhut, Guntram},
  title = {Comprehensive Quantum Chemical Analysis of the Microwave and Infrared Spectra of Cyclopropenone},
  journal = {J. Phys. Chem. A},
  volume = {129},
  pages = {9342--9354},
  year = {2025},
  doi = {10.1021/acs.jpca.5c05429}
}

@article{Conway2021,
  author = {Conway, Eamon K. and Gordon, Iouli E. and Polyansky, Oleg L. and Tennyson, Jonathan},
  title = {Determination of quantum labels based on projections of the total angular momentum on the molecule-fixed axis},
  journal = {J. Quant. Spectrosc. Radiat. Transf.},
  volume = {270},
  pages = {107716},
  year = {2021},
  doi = {10.1016/j.jqsrt.2021.107716}
}

@article{Scribano2008,
  author = {Scribano, Yohann and Benoit, David M.},
  title = {Iterative active-space selection for vibrational configuration interaction calculations using a reduced-coupling {VSCF} basis},
  journal = {Chem. Phys. Lett.},
  volume = {458},
  number = {4-6},
  pages = {384--387},
  year = {2008},
  doi = {10.1016/j.cplett.2008.05.001}
}

@article{Huron1973,
  author = {B. Huron and J. P. Malrieu and P. Rancurel},
  title = {Iterative perturbation calculations of ground and excited state energies from multiconfigurational zeroth-order wavefunctions},
  journal = {J. Chem. Phys.},
  volume = {58},
  pages = {5745},
  year = {1973},
  doi = {10.1063/1.1679199}
}

@article{Buenker1974,
  author = {R. J. Buenker and S. D. Peyerimhoff},
  title = {Individualized configuration selection in {CI} calculations with subsequent energy extrapolation},
  journal = {Theoret. Chim. Acta},
  volume = {35},
  pages = {33},
  year = {1974},
  doi = {10.1007/BF02394557}
}

@book{YurchenkoBook2024,
  author = {S. N. Yurchenko},
  title = {Computational Spectroscopy of Polyatomic Molecules},
  publisher = {CRC Press},
  address = {Boca Raton},
  year = {2024},
  isbn = {978-1-032-22687-3}
}

@article{Rodina2026,
title = {Modeling and analysis of high-resolution H212CO spectra in the range 4450-5000 cm-1},
journal = {Journal of Quantitative Spectroscopy and Radiative Transfer},
volume = {350},
pages = {109757},
year = {2026},
issn = {0022-4073},
doi = {https://doi.org/10.1016/j.jqsrt.2025.109757},
url = {https://www.sciencedirect.com/science/article/pii/S0022407325004194},
author = {A.A. Rodina and A.V. Nikitin and A.E. Protasevich and L. Manceron and M. Rey and Vl.G. Tyuterev}
}

@article{Yachmenev2011,
    author = {Yachmenev, Andrey and Yurchenko, Sergei N. and Jensen, Per and Thiel, Walter},
    title = {A new "spectroscopic" potential energy surface for formaldehyde in its ground electronic state},
    journal = {J. Chem. Phys.},
    volume = {134},
    number = {24},
    pages = {244307},
    year = {2011},
    month = {06},
    issn = {0021-9606},
    xdoi = {10.1063/1.3599927},
    xurl = {https://doi.org/10.1063/1.3599927},
xeprint = {https://pubs.aip.org/aip/jcp/article-pdf/doi/10.1063/1.3599927/15438004/244307\_1\_online.pdf},
}

@article{Nikitin2024,
title = {New quantum assignments and analysis of high-resolution H212CO spectra in the range 3700–4450 cm-1},
journal = {Journal of Quantitative Spectroscopy and Radiative Transfer},
volume = {329},
pages = {109180},
year = {2024},
issn = {0022-4073},
doi = {https://doi.org/10.1016/j.jqsrt.2024.109180},
url = {https://www.sciencedirect.com/science/article/pii/S0022407324002875},
author = {A.V. Nikitin and A.A. Rodina and A.E. Protasevich and L. Manceron and M. Rey and V.G. Tyuterev}
}

@ARTICLE{Refaie015,
  author={Al-Refaie, Ahmed F. and Yachmenev, Andrey and Tennyson, Jonathan and Yurchenko, Sergei N.},
  journal={Monthly Notices of the Royal Astronomical Society}, 
  title={ExoMol line lists - VIII. A variationally computed line list for hot formaldehyde}, 
  year={2015},
  volume={448},
  number={2},
  pages={1704-1714},
  doi={10.1093/mnras/stv091}}

@article{ccf12,
author = {T. B. Adler and G. Knizia and H.-J. Werner},
title = {A simple and efficient {CCSD(T)-F12} approximation},
year = {2007},
journal = "J. Chem. Phys.",
volume = {127},
pages = {221106},
}

@article{n-mode,
author = { Joel M   Bowman  and  Tucker   Carrington  and  Hans-Dieter   Meyer },
title = {Variational quantum approaches for computing vibrational energies of polyatomic molecules},
journal = {Mol. Phys.},
volume = {106},
number = {16-18},
pages = {2145--2182},
year  = {2008},
}

@ARTICLE{MultiLevel,
author={K. Pfl{\"u}ger and M. Paulus and S. Jagiella and T. Burkert and G. Rauhut},
title="Multi-Level Vibrational {SCF} Calculations and {FTIR} Measurements on Furazan",
journal="Theor. Chem. Acc.",
volume=114,
year=2005,
pages="327-332"}

@article{Schriber_JCTC13_2017,
  title     = {Adaptive Configuration Interaction for Computing Challenging Electronic Excited States with Tunable Accuracy},
  volume    = {13},
  issn      = {1549-9626},
  url       = {http://dx.doi.org/10.1021/acs.jctc.7b00725},
  doi       = {10.1021/acs.jctc.7b00725},
  number    = {11},
  journal   = {J. Chem. Theory Comput.},
  publisher = {American Chemical Society (ACS)},
  author    = {Schriber,  Jeffrey B. and Evangelista,  Francesco A.},
  year      = {2017},
  month     = Oct,
  pages     = {5354-5366}
}

@ARTICLE{Avila2012,
        author = {Avila, Gustavo and Carrington, Tucker},
        title = {Solving the vibrational Schr\"odinger equation using bases pruned to include strongly coupled functions and compatible quadratures},
        year = {2012},
        journal = {J. Chem. Phys.},
        volume = {137},
        number = {17},
        doi = {10.1063/1.4764099},
        type = {Article},
}

@Article{Oschetzki2014,
author ="Oschetzki, Dominik and Rauhut, Guntram",
title  ="Pushing the limits in accurate vibrational structure calculations: anharmonic frequencies of lithium fluoride clusters (LiF)$_n${,} $n = 2-10$",
journal  ="Phys. Chem. Chem. Phys.",
year  ="2014",
volume  ="16",
issue  ="31",
pages  ="16426-16435",
publisher  ="The Royal Society of Chemistry"
}

@article{Carbonniere_TCA125_2010,
author={Carbonniere, P. and Dargelos, A. and Pouchan, C.},
title={The VCI-P code: an iterative variation–perturbation scheme for efficient computations of anharmonic vibrational levels and IR intensities of polyatomic molecules},
journal ={Theor. Chem. Acc.},
volume=125,
pages={543-554},
year=2010,
}

@article{christiansen2003,
        Author = {Christiansen, O},
        Journal = {J. Chem. Phys.},
        Pages = {5773-5781},
        Title = {Moller-Plesset perturbation theory for vibrational wave functions},
        Volume = {119},
        Year = {2003},
}

@ARTICLE{Mant2018,
        author = {Mant, Barry P. and Yachmenev, Andrey and Tennyson, Jonathan and Yurchenko, Sergei N.},
        title = {ExoMol molecular line lists - XXVII. Spectra of C2H4},
        year = {2018},
        journal = {MNRAS},
        volume = {478},
        number = {3},
        pages = {3220-3232},
        doi = {10.1093/MNRAS/STY1239},
        type = {Article},
}

@ARTICLE{ml_yagi,
author="K.~Yagi and S.~Hirata and K.~Hirao",
title="Multiresolution potential energy surfaces for vibrational state calculations",
journal="Theor. Chim. Acta",
volume=118,
year=2007,
pages="681-691"}

@ARTICLE{ml_ove,
author="M.~Sparta and M. B.~Hansen and E.~Matito and D.~Toffoli and O.~Christiansen",
title="Using electronic energy derivative information in automated potential energy surface construction for vibrational calculations",
journal="J.~Chem.~Theory Comput.",
volume=6,
year=2010,
pages="3162-3175"}

@article{Rey2022NovelAnalysis,
    title = {{Novel methodology for systematically constructing global effective models from ab initio -based surfaces: A new insight into high-resolution molecular spectra analysis}},
    year = {2022},
    journal = {Journal of Chemical Physics},
    author = {Rey, Michael},
    number = {22},
    volume = {156},
    doi = {10.1063/5.0089097},
    issn = {10897690}
}

@article{Dinu2022,
  author = {Dinu, Dennis F. and Tsch{\"o}pe, Martin and Schr\"oder, Benjamin and Liedl, Klaus R. and Rauhut, Guntram},
  journal = {J. Chem. Phys.},
  number = 15,
  pages = 154107,
  title = {Determination of spectroscopic constants from rovibrational configuration interaction calculations},
  url = {https://doi.org/10.1063/5.0116018},
  volume = 157,
  year = 2022
}

@inbook{RVCI_book_chapter,
author = {Das, Subhasish and Rauhut, Guntram},
publisher = {John Wiley \& Sons, Ltd},
isbn = {9781394474455},
title = {{Automated Rovibrational Configuration Interaction Calculations}},
booktitle = {High‐Resolution Molecular Spectroscopy 1B},
chapter = {1},
pages = {1-29},
doi = {https://doi.org/10.1002/9781394474455.ch1},
year = {2026}
}

@article{Avila2017,
title = {Pruned bases that are compatible with iterative eigensolvers and general potentials: New results for CH3CN},
journal = {Chemical Physics},
volume = {482},
pages = {3-8},
year = {2017},
doi = {https://doi.org/10.1016/j.chemphys.2016.09.023},
author = {Gustavo Avila and Tucker Carrington}
}

@article{Yurchenko2024,
    author = {Yurchenko, Sergei N and Owens, Alec and Kefala, Kyriaki and Tennyson, Jonathan},
    title = {ExoMol line lists – LVII. High accuracy ro-vibrational line list for methane (CH4)},
    journal = {Monthly Notices of the Royal Astronomical Society},
    volume = {528},
    number = {2},
    pages = {3719-3729},
    year = {2024},
    month = {02},
    issn = {0035-8711},
    doi = {10.1093/mnras/stae148},
}

@article{Owens2017,
    author = {Owens, A. and Yachmenev, A. and Thiel, W. and Tennyson, J. and Yurchenko, S. N.},
    title = {ExoMol line lists – XXII. The rotation-vibration spectrum of silane up to 1200 K},
    journal = {Monthly Notices of the Royal Astronomical Society},
    volume = {471},
    number = {4},
    pages = {5025-5032},
    year = {2017},
    month = {11},
    issn = {0035-8711},
    doi = {10.1093/mnras/stx1952},
}

@article{Amrein1985,
author = {A. Amrein and H.-R. Dübal and M. Quack},
title = {Multiple anharmonic resonances in the vibrational overtone spectra of CHClF2 },
journal = {Molecular Physics},
volume = {56},
number = {3},
pages = {727--735},
year = {1985},
publisher = {Taylor \& Francis},
doi = {10.1080/00268978500102671},
}

@article{Krasnoshchekov2013,
    author = {Krasnoshchekov, Sergey V. and Stepanov, Nikolay F.},
    title = {Polyad quantum numbers and multiple resonances in anharmonic vibrational studies of polyatomic molecules},
    journal = {The Journal of Chemical Physics},
    volume = {139},
    number = {18},
    pages = {184101},
    year = {2013},
    month = {11},
    issn = {0021-9606},
    doi = {10.1063/1.4829143},
}

@article{Kellman1990,
    author = {Kellman, Michael E.},
    title = {Approximate constants of motion for vibrational spectra of many‐oscillator systems with multiple anharmonic resonances},
    journal = {The Journal of Chemical Physics},
    volume = {93},
    number = {9},
    pages = {6630-6635},
    year = {1990},
    month = {11},
    issn = {0021-9606},
    doi = {10.1063/1.458930},
}

@article{Yachmenev2013,
    author = {Yachmenev, Andrey and Polyak, Iakov and Thiel, Walter},
    title = {Theoretical rotation-vibration spectrum of thioformaldehyde},
    journal = {The Journal of Chemical Physics},
    volume = {139},
    number = {20},
    pages = {204308},
    year = {2013},
    month = {11},
    issn = {0021-9606},
    doi = {10.1063/1.4832322},
}

@article{SousaSilva2013,
title = {A computed room temperature line list for phosphine},
journal = {Journal of Molecular Spectroscopy},
volume = {288},
pages = {28-37},
year = {2013},
issn = {0022-2852},
doi = {https://doi.org/10.1016/j.jms.2013.04.002},
author = {Clara Sousa-Silva and Sergei N. Yurchenko and Jonathan Tennyson},
}

@article{Avila2019,
    author = {Avila, Gustavo and Mátyus, Edit},
    title = {Toward breaking the curse of dimensionality in (ro)vibrational computations of molecular systems with multiple large-amplitude motions},
    journal = {The Journal of Chemical Physics},
    volume = {150},
    number = {17},
    pages = {174107},
    year = {2019},
    month = {05},
    issn = {0021-9606},
    doi = {10.1063/1.5090846}
}

@article{Sunaga2025,
    author = {Sunaga, Ayaki and Győri, Tibor and Czakó, Gábor and Mátyus, Edit},
    title = {Exact quantum dynamics of methanol: Full-dimensional ab initio potential energy surface of spectroscopic quality and variational vibrational states},
    journal = {The Journal of Chemical Physics},
    volume = {163},
    number = {6},
    pages = {064101},
    year = {2025},
    month = {08},
    issn = {0021-9606},
    doi = {10.1063/5.0283300},
}

@article{Alberto2022,
title = {Variational vibrational states of HCOOH},
journal = {Journal of Molecular Spectroscopy},
volume = {385},
pages = {111617},
year = {2022},
issn = {0022-2852},
doi = {https://doi.org/10.1016/j.jms.2022.111617},
author = {Alberto {Martín Santa Daría} and Gustavo Avila and Edit Mátyus},
}

@article{Matyus2023,
    author = {Mátyus, Edit and Martín Santa Daría, Alberto and Avila, Gustavo},
    title = {Exact quantum dynamics developments for floppy molecular systems and complexes},
    journal = {Chemical Communications},
    volume = {59},
    number = {4},
    pages = {366-381},
    year = {2023},
    month = {01},
    issn = {1359-7345},
    doi = {10.1039/d2cc05123k},
}

@article{AvilaCarrington2009,
    author = {Avila, Gustavo and Carrington, Tucker, Jr.},
    title = {Nonproduct quadrature grids for solving the vibrational Schrödinger equation},
    journal = {The Journal of Chemical Physics},
    volume = {131},
    number = {17},
    pages = {174103},
    year = {2009},
    month = {11},
    issn = {0021-9606},
    doi = {10.1063/1.3246593},
}

@article{AvilaCarrington2011,
    author = {Avila, Gustavo and Carrington, Tucker, Jr.},
    title = {Using a pruned basis, a non-product quadrature grid, and the exact Watson normal-coordinate kinetic energy operator to solve the vibrational Schrödinger equation for C2H4},
    journal = {The Journal of Chemical Physics},
    volume = {135},
    number = {6},
    pages = {064101},
    year = {2011},
    month = {08},
    issn = {0021-9606},
    doi = {10.1063/1.3617249},
}

@article{WangCarrington2001,
    author = {Wang, Xiao-Gang and Carrington, Tucker},
    title = {The Utility of Constraining Basis Function Indices When Using the Lanczos Algorithm to
Calculate Vibrational Energy Levels},
    journal = {The Journal of Physical Chemistry A},
    volume = {105},
    number = {12},
    pages = {2575-2581},
    year = {2001},
    month = {02},
    issn = {1089-5639},
    doi = {10.1021/jp003792s},
}

@Article{Carrington2018Review,
AUTHOR = {Carrington, Tucker},
TITLE = {Iterative Methods for Computing Vibrational Spectra},
JOURNAL = {Mathematics},
VOLUME = {6},
YEAR = {2018},
NUMBER = {1},
ARTICLE-NUMBER = {13},
URL = {https://www.mdpi.com/2227-7390/6/1/13},
ISSN = {2227-7390},
DOI = {10.3390/math6010013}
}

@article{Wang2015,
    author = {Wang, Xiaohong and Carter, Stuart and Bowman, Joel M.},
    title = {Pruning the Hamiltonian Matrix in MULTIMODE: Test
for C2H4 and Application to CH3NO2 Using a New Ab Initio Potential Energy Surface},
    journal = {The Journal of Physical Chemistry A},
    volume = {119},
    number = {47},
    pages = {11632-11640},
    year = {2015},
    month = {11},
    issn = {1089-5639},
    doi = {10.1021/acs.jpca.5b09816},
}

@article{Halonen1983,
    author = {Halonen, L. and Noid, D. W. and Child, M.S.},
    title = {Local mode predictions for excited stretching vibrational states of HCCD and H12C13CH},
    journal = {The Journal of Chemical Physics},
    volume = {78},
    number = {5},
    pages = {2803-2804},
    year = {1983},
    month = {03},
    issn = {0021-9606},
    doi = {10.1063/1.444996},
}

@article{Carter1986,
title = {The variational method for the calculation of RO-vibrational energy levels},
journal = {Computer Physics Reports},
volume = {5},
number = {3},
pages = {117-171},
year = {1986},
issn = {0167-7977},
doi = {https://doi.org/10.1016/0167-7977(86)90006-7},
author = {S. Carter and N.C. Handy},
}

@article{Simmons2023,
    author = {Simmons, Jesse and Carrington, Tucker, Jr.},
    title = {Computing vibrational spectra using a new collocation method with a pruned basis and more points than basis functions: Avoiding quadrature},
    journal = {The Journal of Chemical Physics},
    volume = {158},
    number = {14},
    pages = {144115},
    year = {2023},
    month = {04},
    issn = {0021-9606},
    doi = {10.1063/5.0146703},
}

@article{Kellman1995,
  author  = {Kellman, Michael E.},
  title   = {Algebraic Methods in Spectroscopy},
  journal = {Annual Review of Physical Chemistry},
  year    = {1995},
  volume  = {46},
  pages   = {395--422},
  doi     = {10.1146/annurev.pc.46.100195.002143}
}

@article{BermudezMontana2017,
  author  = {Bermudez-Montaña, Marisol and Lemus, Renato and Pérez-Bernal, Francisco and Carvajal, Miguel},
  title   = {Comprehensive vibrational analysis of {CO$_2$} based on a polyad-preserving model},
  journal = {The European Physical Journal D},
  volume  = {71},
  number  = {6},
  pages   = {147},
  year    = {2017},
  doi     = {10.1140/epjd/e2017-80178-6}
}

@article{Ishikawa1998,
    author = {Ishikawa, Haruki and Nagao, Chioko and Mikami, Naohiko and Field, Robert W.},
    title = {Spectroscopic investigation of the generation of “isomerization” states: Eigenvector analysis of the bend-CP stretch polyad},
    journal = {The Journal of Chemical Physics},
    volume = {109},
    number = {2},
    pages = {492-503},
    year = {1998},
    month = {07},
    issn = {0021-9606},
    doi = {10.1063/1.476586},
}

@article{Barnes2011,
    author = {Barnes, George L. and Kellman, Michael E.},
    title = {Detailed analysis of polyad-breaking spectroscopic Hamiltonians for multiple minima with above barrier motion: Isomerization in HO2},
    journal = {The Journal of Chemical Physics},
    volume = {134},
    number = {7},
    pages = {074108},
    year = {2011},
    month = {02},
    issn = {0021-9606},
    doi = {10.1063/1.3552086},
}

@article{Stampfuss1999,
author = {Stampfuss, P. and Wenzel, W. and Keiter, H.},
title = {The parallel implementation of configuration-selecting multireference configuration interaction method},
journal = {Journal of Computational Chemistry},
volume = {20},
number = {14},
pages = {1559-1570},
doi = {https://doi.org/10.1002/(SICI)1096-987X(19991115)20:14<1559::AID-JCC9>3.0.CO;2-B},
year = {1999}
}

@article{Whitten1969,
    author = {Whitten, J. L. and Hackmeyer, Melvyn},
    title = {Configuration Interaction Studies of Ground and Excited States of Polyatomic Molecules. I. The CI Formulation and Studies of Formaldehyde},
    journal = {The Journal of Chemical Physics},
    volume = {51},
    number = {12},
    pages = {5584-5596},
    year = {1969},
    month = {12},
    issn = {0021-9606},
    doi = {10.1063/1.1671985},
}

@article{Handy2004,
author = {Nicholas C. Handy and Stuart Carter},
title = {Large vibrational variational calculations using ‘multimode’ and an iterative diagonalization technique},
journal = {Molecular Physics},
volume = {102},
number = {21-22},
pages = {2201--2205},
year = {2004},
publisher = {Taylor \& Francis},
doi = {10.1080/00268970410001728870},
}

@article{gamma,
author = { H.-J.   Werner  and  G.   Knizia  and  F. R.   Manby },
xtitle = {Explicitly correlated coupled cluster methods with pair-specific geminals},
journal = {Mol. Phys.},
volume = {109},
xnumber = {3},
pages = {407-417},
year  = {2011},
publisher = {Taylor & Francis},
}

@article{geminals,
author="A. J. May and F. R. Manby", 
journal="J. Chem. Phys.",
xtitle="An explicitly correlated second order {M}{\o}ller-{P}lesset theory using a frozen {G}aussian geminal",
volume=121, 
pages=4479,
year=2004}

@article{Goodson,
    author = "David Z. Goodson",
    xtitle = "Extrapolating the coupled-cluster sequence toward the full configuration-interaction limit",
    journal = "J. Chem. Phys.",
    year = 2002,
    volume= 116,
    pages = {6948-6956},
    doi = {10.1093/mnras/stad1248},
}

@article{h2co_exp,   
journal = {J. Mol. Spectrosc.},
volume = {221},
number = {2},
pages = {192-198},
year = {2003},
issn = {0022-2852},
doi = {https://doi.org/10.1016/S0022-2852(03)00207-8},
url = {https://www.sciencedirect.com/science/article/pii/S0022285203002078},
author = {A. Perrin and F. Keller and J.-M. Flaud},
}

@article{h2co_exp_2,
journal = {J. Mol. Struct.},
volume = {780-781},
pages = {28-44},
year = {2006},
xnote = {Spectroscopic and Theoretical Determination of Molecular Properties},
issn = {0022-2860},
doi = {https://doi.org/10.1016/j.molstruc.2005.03.052},
url = {https://www.sciencedirect.com/science/article/pii/S0022286005005533},
author = {A. Perrin and A. Valentin and L. Daumont},
}

@article{h2ccs_exp_1,
xtitle = {Resonance between the Ground Vibrational State and the Lowest Frequency Bending Mode of Thioketene},
journal = {J. Mol. Spectrosc.},
volume = {175},
number = {2},
pages = {377-385},
year = {1996},
issn = {0022-2852},
doi = {https://doi.org/10.1006/jmsp.1996.0043},
url = {https://www.sciencedirect.com/science/article/pii/S0022285296900430},
author = {D. McNaughton and E.G. Robertson and L.D. Hatherley},
}

@article{h2ccs_exp_2,
xtitle = {Fourier transform infrared spectrum of thioketene, H2CCS},
journal = {J. Mol. Spectrosc.},
volume = {114},
number = {2},
pages = {473-482},
year = {1985},
issn = {0022-2852},
doi = {https://doi.org/10.1016/0022-2852(85)90239-5},
url = {https://www.sciencedirect.com/science/article/pii/0022285285902395},
author = {H.W. Kroto and D. McNaughton},
}

\end{document}


\title[]{\begin{center} Supplementary Material for:\end{center} Basis pruning in rovibrational configuration interaction (RVCI) calculations\\}

\author{Kemal Oenen}
\affiliation{Institute for Theoretical Chemistry, University of Stuttgart, Pfaffenwaldring 55, 70569 Stuttgart, Germany}%
\affiliation{Institute for Quantum Electronics, Department of Physics, ETH Zürich, Otto-Stern-Weg 1, 8093 Zürich, Switzerland}

\author{Subhasish Das}
\affiliation{Institute for Theoretical Chemistry, University of Stuttgart, Pfaffenwaldring 55, 70569 Stuttgart, Germany}%

\author{Guntram Rauhut$^{*,}$}
\email[Author to whom correspondence should be addressed: ]{rauhut@theochem.uni-stuttgart.de}
\affiliation{Institute for Theoretical Chemistry, University of Stuttgart, Pfaffenwaldring 55, 70569 Stuttgart, Germany}%

\date{\today}

\maketitle

\section{Formaldehyde (H$_2$CO)}

\subsection{Computational details}

\begin{table}[t]
\begin{center}
\caption{Scaling factors for the one-mode terms of the PES of formaldehyde.\label{tab:scalfac}}
\begin{tabular}{cc}
\hline \hline
Mode$^a$ & Factor \\
\hline
1 & 1.00072719 \\
2 & 1.00152754 \\
3 & 1.00049701 \\
4 & 0.99934170 \\
5 & 1.00049052 \\
6 & 1.00004532 \\
\hline \hline
\end{tabular}
\end{center}
{\footnotesize
a) Modes are ordered by increasing fundamental frequency, from mode 1 (lowest) to mode 6 (highest, C--H stretch).}
\end{table}

An $n$-mode expansion of the potential energy surface (PES), including up to four-mode coupling terms, was obtained using all-electron explicitly correlated coupled-cluster theory, i.e., CCSD(T)-F12b/cc-pCVTZ-F12. As recommended, different exponents of 0.8, 1.7, and 2.2 were used for the Slater geminal functions of the core-core, core-valence, and valence-valence orbital pairs, respectively \cite{geminals,gamma}. High-order coupled-cluster contributions were approximated using Goodson's continued-fraction approach \cite{Goodson}. The PES is provided in the {\sc Zenodo} database via the link given in the Data Availability Statement. The grid-based PES was transformed into an analytical representation using nine monomials per mode. The one-mode terms were subsequently scaled by the factors listed in Table~\ref{tab:scalfac}. These factors were determined by a least-squares fit of the VCI frequencies to experimental reference data. The resulting PES was employed in all subsequent VCI and RVCI calculations. 

The VCI calculations used a basis of Hartree products constructed from VSCF ground-state modals expanded in 18 distributed Gaussians. The correlation space included excitations up to sextuples, with a maximum excitation of nine quanta in an individual mode and a maximum sum of vibrational quantum numbers of 18. Vibrational angular momentum terms up to first order were included in the diagonal elements of the VCI matrix, while a constant $\mu$-tensor was used for the off-diagonal elements. The resulting VCI fundamental frequencies are listed in Table~\ref{tab:freq_h2co} together with the corresponding experimental values, infrared intensities, and assignments used in the main text. The conventional spectroscopic labels $\nu_i$ denote fundamental transitions from the vibrational ground state to the first excited state of the respective mode. For the assignment of rovibrational states, we instead refer to the corresponding upper vibrational states as $i^1$, with the mode index $i$ assigned in order of increasing fundamental frequency.

\begin{table}
\caption{Experimental and VCI fundamental frequencies (in cm$^{-1}$) and infrared intensities (in km/mol) of formaldehyde.\label{tab:freq_h2co}}
\begin{center}
\begin{tabular}{cccccr}
    \hline \hline
    State$^a$ & Trans.$^b$ & Sym. & Exp.$^c$ & VCI & IR \\
    \hline
    $1^1$ & $\nu_6$ & B$_2$ & 1167.26 & 1167.2 &   5.2 \\
    $2^1$ & $\nu_5$ & B$_1$ & 1249.10 & 1249.5 &  20.0 \\
    $3^1$ & $\nu_3$ & A$_1$ & 1500.18 & 1500.0 &  26.2 \\
    $4^1$ & $\nu_2$ & A$_1$ & 1746.01 & 1746.1 & 150.5 \\
    $5^1$ & $\nu_1$ & A$_1$ & 2782.46 & 2782.3 &  54.6 \\
    $6^1$ & $\nu_4$ & B$_1$ & 2843.33 & 2843.4 &  75.6 \\
    \hline \hline
\end{tabular}
\end{center}
{\footnotesize
a) Labeling of the upper vibrational states in order of increasing fundamental frequency.\\
b) Labeling of the fundamental transitions according to the spectroscopic convention based on the irreducible representations.\\
c) Experimental data taken from Refs.\ \onlinecite{h2co_exp,h2co_exp_2}.}
\end{table}

For all RVCI calculations, a maximum rotational quantum number of $J_{\max}=60$ was employed. The $\mu$-tensor expansions entering the rotational and Coriolis-coupling terms were truncated after second and first order, respectively. The rovibrational spectra presented in the SM were calculated at $T=300$~K using the separability approximation $Q=Q_{\mathrm{vib}}Q_{\mathrm{rot}}$ for the partition function. 

Finally, to complement the description of the basis-pruning calculations in the main text, we report here the numbers of rovibrational eigenvalues monitored during the individual pruning steps. During preselection, the numbers of monitored eigenvalues across windows 1--3 were 3370, 3567, and 3519. In the main pruning loop, they were 363, 121, and 242.

\subsection{Assignment and matching statistics}
For the state-energy comparisons, rovibrational levels were first filtered according to the assignment-confidence criterion $A\geq 0.5$. Ambiguous duplicate assignments were subsequently removed before states obtained with the reference, polyad-truncated, and pruned bases were matched according to their quantum numbers. For the transition-frequency analysis, duplicate transition assignments were removed before matching the remaining transitions. The complete filtering and matching statistics are summarized in Tables~\ref{tab:h2co_state_matching_j10} and \ref{tab:h2co_state_matching} for the state energies with $J\leq 10$ and $J\leq 60$, respectively, and in Table~\ref{tab:h2co_transition_matching} for the transition frequencies.

\begin{table}[t]
\caption{Statistics of assignment-confidence filtering and state matching for the H$_2$CO rovibrational energies with $J\leq10$. $N_{\mathrm{ref}}$ and $N_{\mathrm{red}}$ denote the numbers of levels obtained with the reference and reduced vibrational bases, respectively, with superscripts ``all'' and $A$ referring to all states and those remaining after application of the assignment-confidence criterion $A\geq0.5$. $N_{\mathrm{match}}$ gives the number of uniquely matched levels. Duplicate removal is not shown because it does not alter the counts.}
\label{tab:h2co_state_matching_j10}
\centering
\begin{tabular}{llrrrrr}
\hline\hline
Vib.\ assign.\ & Basis
& $N_{\mathrm{ref}}^{\mathrm{all}}$
& $N_{\mathrm{ref}}^{A}$
& $N_{\mathrm{red}}^{\mathrm{all}}$
& $N_{\mathrm{red}}^{A}$
& $N_{\mathrm{match}}$ \\
\hline

$1^1$ & $P_1$    & 121 & 121 & 121 & 121 & 121 \\
      & $P_2$    & 121 & 121 & 121 & 121 & 121 \\
      & $P_3$    & 121 & 121 & 121 & 121 & 121 \\
      & Win.\ 1  & 121 & 121 & 121 & 121 & 121 \\[2pt]
$2^1$ & $P_1$    & 121 & 121 & 121 & 121 & 121 \\
      & $P_2$    & 121 & 121 & 121 & 121 & 121 \\
      & $P_3$    & 121 & 121 & 121 & 121 & 121 \\
      & Win.\ 1  & 121 & 121 & 121 & 121 & 121 \\[2pt]
$3^1$ & $P_1$    & 121 & 121 & 121 & 121 & 121 \\
      & $P_2$    & 121 & 121 & 121 & 121 & 121 \\
      & $P_3$    & 121 & 121 & 121 & 121 & 121 \\
      & Win.\ 1  & 121 & 121 & 121 & 121 & 121 \\[2pt]
$4^1$ & $P_1$    & 121 & 121 & 121 & 121 & 121 \\
      & $P_2$    & 121 & 121 & 121 & 121 & 121 \\
      & $P_3$    & 121 & 121 & 121 & 121 & 121 \\
      & Win.\ 2  & 121 & 121 & 121 & 121 & 121 \\[2pt]
$5^1$ & $P_2$    & 121 & 121 & 121 & 121 & 121 \\
      & $P_3$    & 121 & 121 & 121 & 121 & 121 \\
      & Win.\ 3  & 121 & 121 & 121 & 121 & 121 \\[2pt]
$6^1$ & $P_2$    & 123 & 117 & 121 & 115 & 115 \\
      & $P_3$    & 123 & 117 & 125 & 117 & 117 \\
      & Win.\ 3  & 123 & 117 & 121 & 115 & 115 \\
\hline\hline
\end{tabular}
\end{table}

\begin{table}[t]
\caption{Same as Table~\ref{tab:h2co_state_matching_j10}, but for the \ch{H2CO} rovibrational energies with $J\leq60$.}ett
\label{tab:h2co_state_matching}
\centering
\begin{tabular}{llrrrrr}
\hline\hline
Vib.\ assign.\ & Basis
& $N_{\mathrm{ref}}^{\mathrm{all}}$
& $N_{\mathrm{ref}}^{A}$
& $N_{\mathrm{red}}^{\mathrm{all}}$
& $N_{\mathrm{red}}^{A}$
& $N_{\mathrm{match}}$ \\
\hline
$1^1$ & $P_1$    & 3981 & 2507 & 3695 & 3318 & 2455 \\
      & $P_2$    & 3981 & 2507 & 3706 & 2676 & 2481 \\
      & $P_3$    & 3981 & 2507 & 3703 & 2680 & 2499 \\
      & Win.\ 1  & 3981 & 2507 & 3734 & 2657 & 2463 \\[2pt]
$2^1$ & $P_1$    & 3698 & 2100 & 3628 & 2871 & 1926 \\
      & $P_2$    & 3698 & 2100 & 3712 & 2624 & 2016 \\
      & $P_3$    & 3698 & 2100 & 3674 & 2392 & 2082 \\
      & Win.\ 1  & 3698 & 2100 & 3681 & 2389 & 2072 \\[2pt]
$3^1$ & $P_1$    & 3711 & 2707 & 3717 & 2653 & 2410 \\
      & $P_2$    & 3711 & 2707 & 3912 & 2835 & 2687 \\
      & $P_3$    & 3711 & 2707 & 3501 & 2734 & 2704 \\
      & Win.\ 1  & 3711 & 2707 & 3742 & 2876 & 2684 \\[2pt]
$4^1$ & $P_1$    & 3786 & 2958 & 3840 & 3471 & 2894 \\
      & $P_2$    & 3786 & 2958 & 3799 & 3523 & 2947 \\
      & $P_3$    & 3786 & 2958 & 3963 & 2900 & 2854 \\
      & Win.\ 2  & 3786 & 2958 & 3800 & 3226 & 2696 \\[2pt]
$5^1$ & $P_2$    & 4015 & 2589 & 4005 & 2825 & 2452 \\
      & $P_3$    & 4015 & 2589 & 4074 & 2424 & 2313 \\
      & Win.\ 3  & 4015 & 2589 & 3821 & 2096 & 1976 \\[2pt]
$6^1$ & $P_2$    & 3802 & 667  & 4741 & 889  & 559  \\
      & $P_3$    & 3802 & 667  & 4213 & 761  & 636  \\
      & Win.\ 3  & 3802 & 667  & 4235 & 1198 & 564  \\
\hline\hline
\end{tabular}
\end{table}

\begin{table}[t]
\centering
\caption{Statistics of duplicate removal and transition matching for the H$_2$CO rovibrational frequencies with $J\leq60$. $N_{\mathrm{ref}}$ and $N_{\mathrm{red}}$ denote the numbers of transitions obtained with the reference and reduced vibrational bases, respectively, with superscripts ``all'' and ``d'' referring to all transitions and those remaining after removal of ambiguous duplicate transition assignments. $N_{\mathrm{match}}$ gives the number of uniquely matched transitions.}
\label{tab:h2co_transition_matching}
\begin{tabular}{llrrrrr}
\hline\hline
Fund.\ band & Basis
& $N_{\mathrm{ref}}^{\mathrm{all}}$
& $N_{\mathrm{ref}}^{\mathrm{d}}$
& $N_{\mathrm{red}}^{\mathrm{all}}$
& $N_{\mathrm{red}}^{\mathrm{d}}$
& $N_{\mathrm{match}}$ \\
\hline
$1^1$ & $P_1$    & 7976 & 7976 & 7474 & 7474 & 6569 \\
      & $P_2$    & 7976 & 7976 & 7614 & 7614 & 7066 \\
      & $P_3$    & 7976 & 7976 & 7814 & 7814 & 7321 \\
      & Win.\ 1  & 7976 & 7976 & 7554 & 7554 & 7296 \\[2pt]
$2^1$ & $P_1$    & 8174 & 8100 & 7290 & 7244 & 6710 \\
      & $P_2$    & 8174 & 8100 & 7992 & 7946 & 7450 \\
      & $P_3$    & 8174 & 8100 & 8231 & 8183 & 7752 \\
      & Win.\ 1  & 8174 & 8100 & 8095 & 8021 & 7737 \\[2pt]
$3^1$ & $P_1$    & 7642 & 7610 & 7281 & 6497 & 5776 \\
      & $P_2$    & 7642 & 7610 & 7356 & 7352 & 6658 \\
      & $P_3$    & 7642 & 7610 & 7594 & 7560 & 7063 \\
      & Win.\ 1  & 7642 & 7610 & 7421 & 7417 & 7064 \\[2pt]
$4^1$ & $P_1$    & 6804 & 6804 & 6431 & 6431 & 5803 \\
      & $P_2$    & 6804 & 6804 & 6417 & 6375 & 6092 \\
      & $P_3$    & 6804 & 6804 & 6605 & 6569 & 6313 \\
      & Win.\ 2  & 6804 & 6804 & 6682 & 6354 & 5843 \\[2pt]
$5^1$ & $P_2$    & 7632 & 7283 & 7189 & 6851 & 5622 \\
      & $P_3$    & 7632 & 7283 & 7564 & 7168 & 6164 \\
      & Win.\ 3  & 7632 & 7283 & 7498 & 7300 & 5715 \\[2pt]
$6^1$ & $P_2$    & 8003 & 7225 & 7920 & 6120 & 5064 \\
      & $P_3$    & 8003 & 7225 & 7779 & 6981 & 6055 \\
      & Win.\ 3  & 8003 & 7225 & 8083 & 6798 & 5399 \\
\hline\hline
\end{tabular}
\end{table}

\subsection{RVCI spectra}
\begin{figure*}[p]
    \centering
    \includegraphics[width=0.99\textwidth]{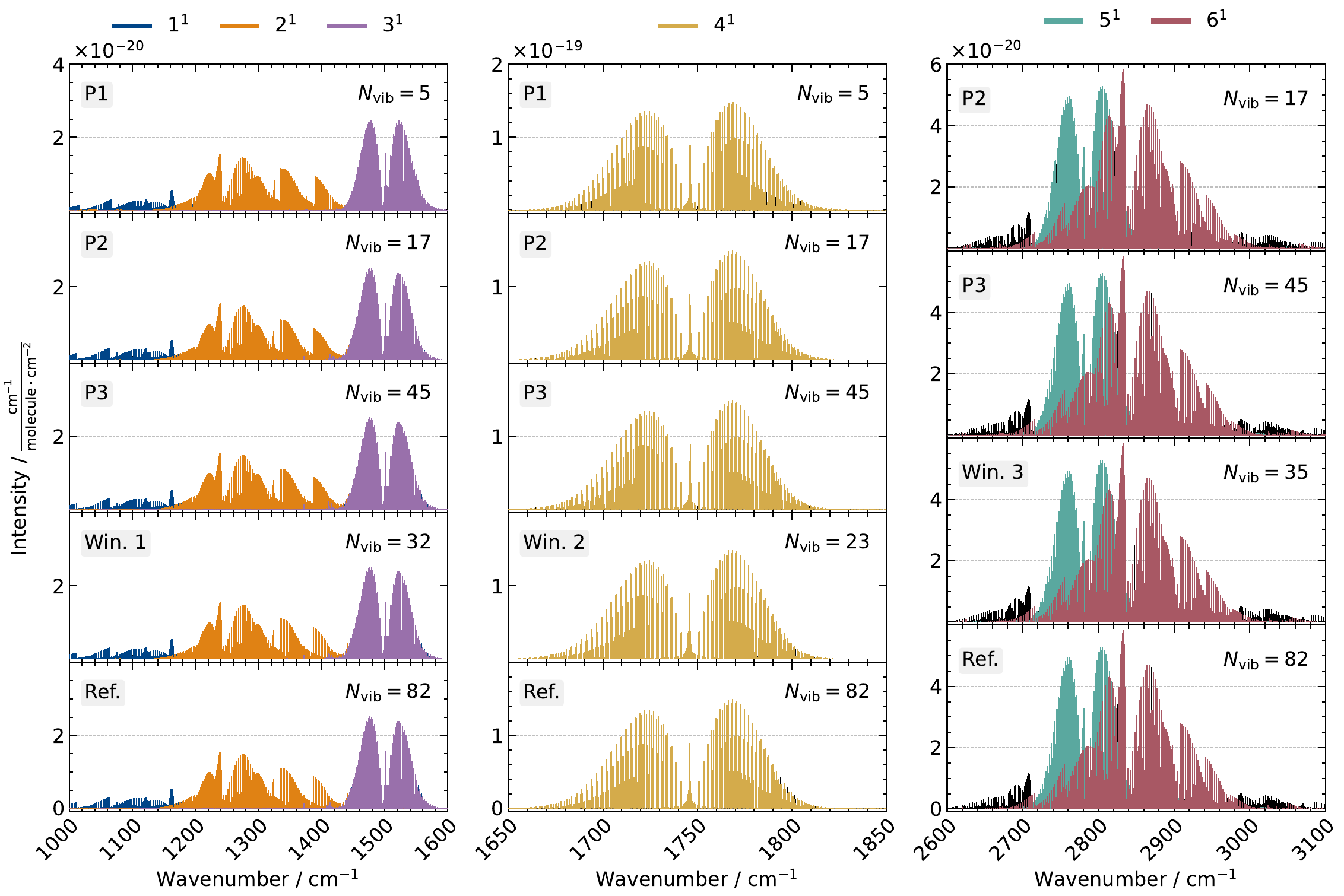}
    \caption{Calculated \ch{H2CO} RVCI stick spectra at $T=300$~K for windows 1--3. The complete spectra are shown in black, with transitions whose upper states are assigned to the respective target fundamentals highlighted in color. Results are shown for the polyad-based and pruned vibrational bases and the 82-state reference calculation.}
    \label{fig:h2co_spectra}
\end{figure*}

Figure~\ref{fig:h2co_spectra} shows the calculated RVCI stick spectra of \ch{H2CO} obtained with the pruned and polyad-based vibrational bases, together with the 82-state reference calculation. As discussed in the main text, the spectra are virtually indistinguishable on the scale of the complete bands.
\section{Thioketene (H$_2$CCS)}

\subsection{Computational details}

An $n$-mode expansion truncated after 4th order has been used to represent the potential energy surface. A multi-level scheme has been employed, in which the 1D and 2D terms were determined at the CCSD(T)-F12b/cc-pVTZ-F12 level including core correlation and approximate correction for high-order coupled-cluster terms, i.e., Goodson's continued fraction approach. The 3D terms employed the explicitly correlated distinguishable clusters approximation, DCSD-F12b/cc-pVDZ-F12 of Kats and Manby and the 4D terms were evaluated at the MP2/cc-pVDZ level. The subsequent VCI calculations include up to sextuple excitation and a maximal sum of quantum numbers of 15. Vibrational angular momentum terms include a $\mu$-tensor truncated after 0th order. Results for the fundamental vibrational frequencies are summarized in Table \ref{tab:supp1}. Experimental data were taken from the compilation provided in the NIST database. Gas phase values have been used for comparison throughout. Note that the deviation of the VCI results from the experimental data is largest when no decimals have been provided to the experimental value. For all RVCI calculations, $J_{\max}=110$ was employed, while all other settings and the treatment of the rovibrational spectra followed those detailed for \ch{H2CO}.

\begin{table}
\caption{Experimental and VCI fundamental frequencies (in cm$^{-1}$) and infrared intensities (in km/mol) of thioketene.\label{tab:supp1}}
\begin{center}
    \begin{tabular}{llld{5.3}d{5.3}d{5.3}}
        \hline 
        State$^a$ & Trans.$^b$ & Sym. & \multicolumn{1}{c}{Exp.$^c$} & \multicolumn{1}{c}{VCI} & \multicolumn{1}{c}{IR} \\
        \hline
        $1^1$ & $\nu_7$ & B$_1$ &  357.45 &  356.4  &   2.6  \\
        $2^1$ & $\nu_9$ & B$_2$ &  413.85 &  414.9  &   6.5  \\
        $3^1$ & $\nu_8$ & B$_2$ &  698.09 &  698.2  &  68.6  \\
        $4^1$ & $\nu_4$ & A$_1$ &  850    &  861.5  &  11.5  \\
        $5^1$ & $\nu_6$ & B$_1$ &  921.6  &  921.1  &   1.1  \\
        $6^1$ & $\nu_3$ & A$_1$ & 1331.67 & 1331.7  &   4.6  \\
        $7^1$ & $\nu_2$ & A$_1$ & 1757    & 1763.6  & 288.8  \\
        $8^1$ & $\nu_1$ & A$_1$ & 3020    & 3025.5  &   4.2  \\
        $9^1$ & $\nu_5$ & B$_1$ & 3107.33 & 3107.2  &   1.5  \\
        \hline
    \end{tabular}
\end{center}
{\footnotesize
a) Labeling of the upper vibrational states in order of increasing fundamental frequency.\\
b) Labeling of the fundamental transitions according to the spectroscopic convention based on the irreducible representations.\\
c) Experimental data taken from Refs.\ \onlinecite{h2ccs_exp_1,h2ccs_exp_2}.}
\end{table}

To conclude, we provide some additional details on the basis-pruning calculations. For the low-frequency spectral region containing the $1^1$--$5^1$ fundamentals, 12321 eigenvalues were monitored during the preselection and 599 during the main pruning loop. For the calculations targeting the $7^1$ fundamental, the corresponding numbers were 10757 and 121, respectively. Starting from the ground state and the $7^1$ fundamental, the preselection added 13 states, followed by 11 states during two iterations of the main pruning loop, yielding a final basis of 26 states. Without preselection, 14 states were added during three iterations, yielding a final basis of 16 states.

\subsection{Assignment and matching statistics}

The assignment and matching analysis described above for H$_2$CO was likewise carried out for all five monitored H$_2$CCS fundamentals at each of the investigated screening thresholds $Q_3^{\rm scr}$. The resulting statistics for the rovibrational energies with $J\leq110$ are reported in Table~\ref{tab:h2ccs_state_matching}, while those for the transition frequencies are given in Table~\ref{tab:h2ccs_transition_matching}.

\begin{table}[t]
\centering
\caption{Statistics of assignment-confidence filtering, duplicate removal, and state matching for the H$_2$CCS rovibrational energies with $J\leq110$. $N_{\mathrm{ref}}$ and $N_{\mathrm{red}}$ denote the numbers of levels obtained with the reference and reduced vibrational bases, respectively, with superscripts ``all'', $A$, and $A,\mathrm{d}$ referring to all states, those satisfying the assignment-confidence criterion $A\geq0.5$, and those remaining after subsequent removal of ambiguous duplicate assignments. $N_{\mathrm{match}}$ gives the number of uniquely matched levels.}
\label{tab:h2ccs_state_matching}
\begin{tabular}{llrrrrrrr}
\hline\hline
Vib.\ assign.\ & $Q_3^{\rm scr}$
& $N_{\rm ref}^{\rm all}$
& $N_{\rm ref}^{A}$
& $N_{\rm ref}^{A,\mathrm{d}}$
& $N_{\rm red}^{\rm all}$
& $N_{\rm red}^{A}$
& $N_{\rm red}^{A,\mathrm{d}}$
& $N_{\rm match}$ \\
\hline
$1^1$ & $10^{-1}$ & 12336 & 1962 & 1962 & 9835  & 2585 & 2585 & 1958 \\
      & $10^{-2}$ & 12336 & 1962 & 1962 & 11375 & 2683 & 2683 & 1962 \\
      & $10^{-3}$ & 12336 & 1962 & 1962 & 10815 & 2007 & 2007 & 1958 \\
      & $10^{-4}$ & 12336 & 1962 & 1962 & 12420 & 1976 & 1976 & 1950 \\
      & $10^{-5}$ & 12336 & 1962 & 1962 & 9504  & 1968 & 1968 & 1952 \\[2pt]

$2^1$ & $10^{-1}$ & 11627 & 1861 & 1861 & 12333 & 3789 & 3789 & 1805 \\
      & $10^{-2}$ & 11627 & 1861 & 1861 & 7439  & 1899 & 1899 & 1837 \\
      & $10^{-3}$ & 11627 & 1861 & 1861 & 10277 & 1927 & 1927 & 1851 \\
      & $10^{-4}$ & 11627 & 1861 & 1861 & 12853 & 1862 & 1862 & 1858 \\
      & $10^{-5}$ & 11627 & 1861 & 1861 & 12668 & 1863 & 1863 & 1859 \\[2pt]

$3^1$ & $10^{-1}$ & 15875 & 3351 & 3351 & 12793 & 9431 & 9431 & 3345 \\
      & $10^{-2}$ & 15875 & 3351 & 3351 & 14341 & 3745 & 3745 & 2711 \\
      & $10^{-3}$ & 15875 & 3351 & 3351 & 16857 & 2636 & 2636 & 2538 \\
      & $10^{-4}$ & 15875 & 3351 & 3351 & 12273 & 3463 & 3463 & 3217 \\
      & $10^{-5}$ & 15875 & 3351 & 3351 & 13724 & 3365 & 3365 & 3279 \\[2pt]

$4^1$ & $10^{-1}$ & 12905 & 2468 & 2468 & 12543 & 9143 & 9143 & 2464 \\
      & $10^{-2}$ & 12905 & 2468 & 2468 & 17243 & 8696 & 8696 & 2350 \\
      & $10^{-3}$ & 12905 & 2468 & 2468 & 17676 & 3049 & 3027 & 2357 \\
      & $10^{-4}$ & 12905 & 2468 & 2468 & 16978 & 3551 & 3529 & 2334 \\
      & $10^{-5}$ & 12905 & 2468 & 2468 & 12293 & 2842 & 2834 & 2444 \\[2pt]

$5^1$ & $10^{-1}$ & 11798 & 3749 & 3749 & 12877 & 7631 & 7631 & 3149 \\
      & $10^{-2}$ & 11798 & 3749 & 3749 & 9823  & 3365 & 3365 & 2933 \\
      & $10^{-3}$ & 11798 & 3749 & 3749 & 14380 & 3354 & 3354 & 2934 \\
      & $10^{-4}$ & 11798 & 3749 & 3749 & 13490 & 4398 & 4398 & 3568 \\
      & $10^{-5}$ & 11798 & 3749 & 3749 & 14939 & 3929 & 3929 & 3567 \\
\hline\hline
\end{tabular}
\end{table}

\begin{table}[t]
\centering
\caption{Statistics of duplicate removal and transition matching for the H$_2$CCS rovibrational frequencies with $J\leq110$. $N_{\mathrm{ref}}$ and $N_{\mathrm{red}}$ denote the numbers of transitions obtained with the reference and reduced vibrational bases, respectively, with superscripts ``all'' and ``d'' referring to all transitions and those remaining after removal of ambiguous duplicate transition assignments. $N_{\mathrm{match}}$ gives the number of uniquely matched transitions.}
\label{tab:h2ccs_transition_matching}
\begin{tabular}{llrrrrr}
\hline\hline
Fund.\ band & $Q_3^{\rm scr}$
& $N_{\rm ref}^{\rm all}$
& $N_{\rm ref}^{\mathrm{d}}$
& $N_{\rm red}^{\rm all}$
& $N_{\rm red}^{\mathrm{d}}$
& $N_{\rm match}$ \\
\hline
$1^1$ & $10^{-1}$ & 18584 & 18528 & 17701 & 17629 & 14458 \\
      & $10^{-2}$ & 18584 & 18528 & 17591 & 17567 & 14538 \\
      & $10^{-3}$ & 18584 & 18528 & 16551 & 16479 & 14492 \\
      & $10^{-4}$ & 18584 & 18528 & 16994 & 16923 & 16230 \\
      & $10^{-5}$ & 18584 & 18528 & 17594 & 17531 & 16933 \\[2pt]
$2^1$ & $10^{-1}$ & 14449 & 14449 & 15992 & 15992 & 12596 \\
      & $10^{-2}$ & 14449 & 14449 & 16553 & 16553 & 13310 \\
      & $10^{-3}$ & 14449 & 14449 & 14121 & 14121 & 13085 \\
      & $10^{-4}$ & 14449 & 14449 & 13797 & 13797 & 13264 \\
      & $10^{-5}$ & 14449 & 14449 & 14288 & 14288 & 13600 \\[2pt]
$3^1$ & $10^{-1}$ & 20332 & 19700 & 19169 & 19049 & 16826 \\
      & $10^{-2}$ & 20332 & 19700 & 19601 & 19601 & 16921 \\
      & $10^{-3}$ & 20332 & 19700 & 19785 & 19419 & 17655 \\
      & $10^{-4}$ & 20332 & 19700 & 19610 & 19024 & 17733 \\
      & $10^{-5}$ & 20332 & 19700 & 19892 & 19368 & 18041 \\[2pt]
$4^1$ & $10^{-1}$ & 20273 & 18191 & 16776 & 16752 & 14061 \\
      & $10^{-2}$ & 20273 & 18191 & 18058 & 17500 & 14335 \\
      & $10^{-3}$ & 20273 & 18191 & 18826 & 17646 & 15424 \\
      & $10^{-4}$ & 20273 & 18191 & 19225 & 17315 & 15232 \\
      & $10^{-5}$ & 20273 & 18191 & 19195 & 17543 & 15997 \\[2pt]
$5^1$ & $10^{-1}$ & 21336 & 19212 & 19763 & 19667 & 16640 \\
      & $10^{-2}$ & 21336 & 19212 & 19861 & 19617 & 16479 \\
      & $10^{-3}$ & 21336 & 19212 & 19928 & 18274 & 15240 \\
      & $10^{-4}$ & 21336 & 19212 & 19679 & 18323 & 16305 \\
      & $10^{-5}$ & 21336 & 19212 & 19841 & 18365 & 16767 \\
\hline\hline
\end{tabular}
\end{table} 

\subsection{State-energy deviation maps}
\begin{figure*}[p]
\centering
\includegraphics[width=\textwidth]{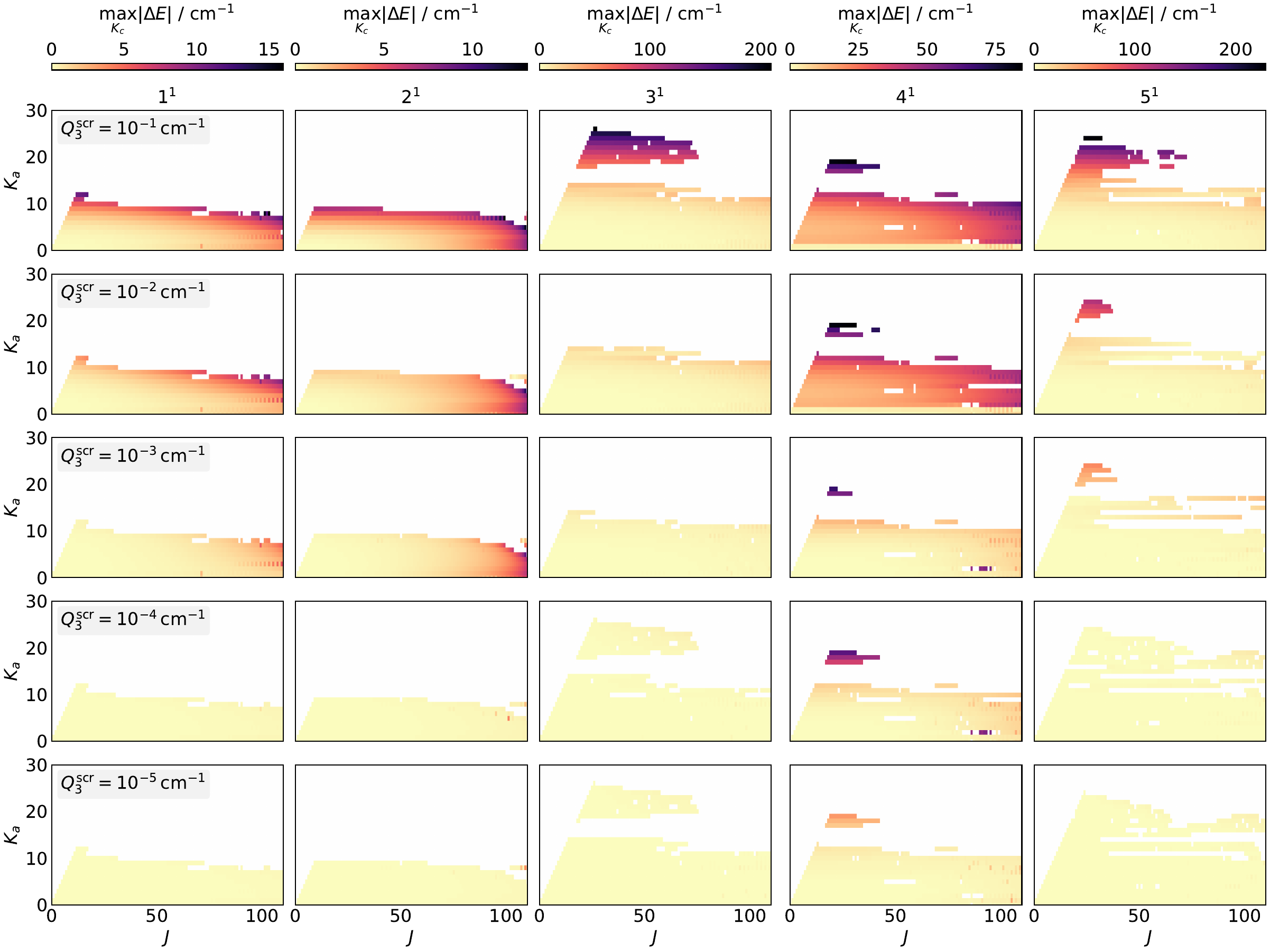}
\caption{Absolute differences in the H$_2$CCS rovibrational state energies between the pruned and 102-state reference calculations, mapped onto the $(J,K_a)$ plane for the five monitored fundamentals $1^1$--$5^1$ (columns) and screening thresholds $Q_3^{\rm scr}=10^{-1}$--$10^{-5}\,\mathrm{cm}^{-1}$ (rows). For each $(J,K_a)$ combination, the maximum absolute deviation over the matched $K_c$ levels is shown. White regions indicate levels excluded by the $A\geq0.5$ assignment-confidence criterion or lacking a unique match to the reference. Note that different color scales are used for the individual fundamentals.}
\label{fig:h2ccs_energy_heatmaps}
\end{figure*}

To complement the state-energy deviations discussed in the main text, Figure~\ref{fig:h2ccs_energy_heatmaps} resolves the differences from the 102-state reference with respect to the rotational quantum numbers $J$ and $K_a$ for all five monitored H$_2$CCS fundamentals and all investigated screening thresholds $Q_3^{\rm scr}$. For each $(J,K_a)$ combination, the maximum absolute deviation over the matched $K_c$ levels is shown. Only reliably assigned and uniquely matched levels enter the comparison, such that unoccupied regions correspond to $(J,K_a)$ pairs for which no matched level remains after filtering. The resulting sparsity is particularly pronounced for high-$K_a$ levels with $K_a$ approaching $J$, where increasingly ambiguous assignments limit the number of levels compared. At the same time, the maps illustrate how tightening $Q_3^{\rm scr}$ progressively improves the agreement with the reference, including in regions exhibiting the largest energy deviations at looser thresholds.

\subsection{RVCI spectra}
\begin{figure*}[p]
    \centering
    \includegraphics[width=0.99\textwidth]{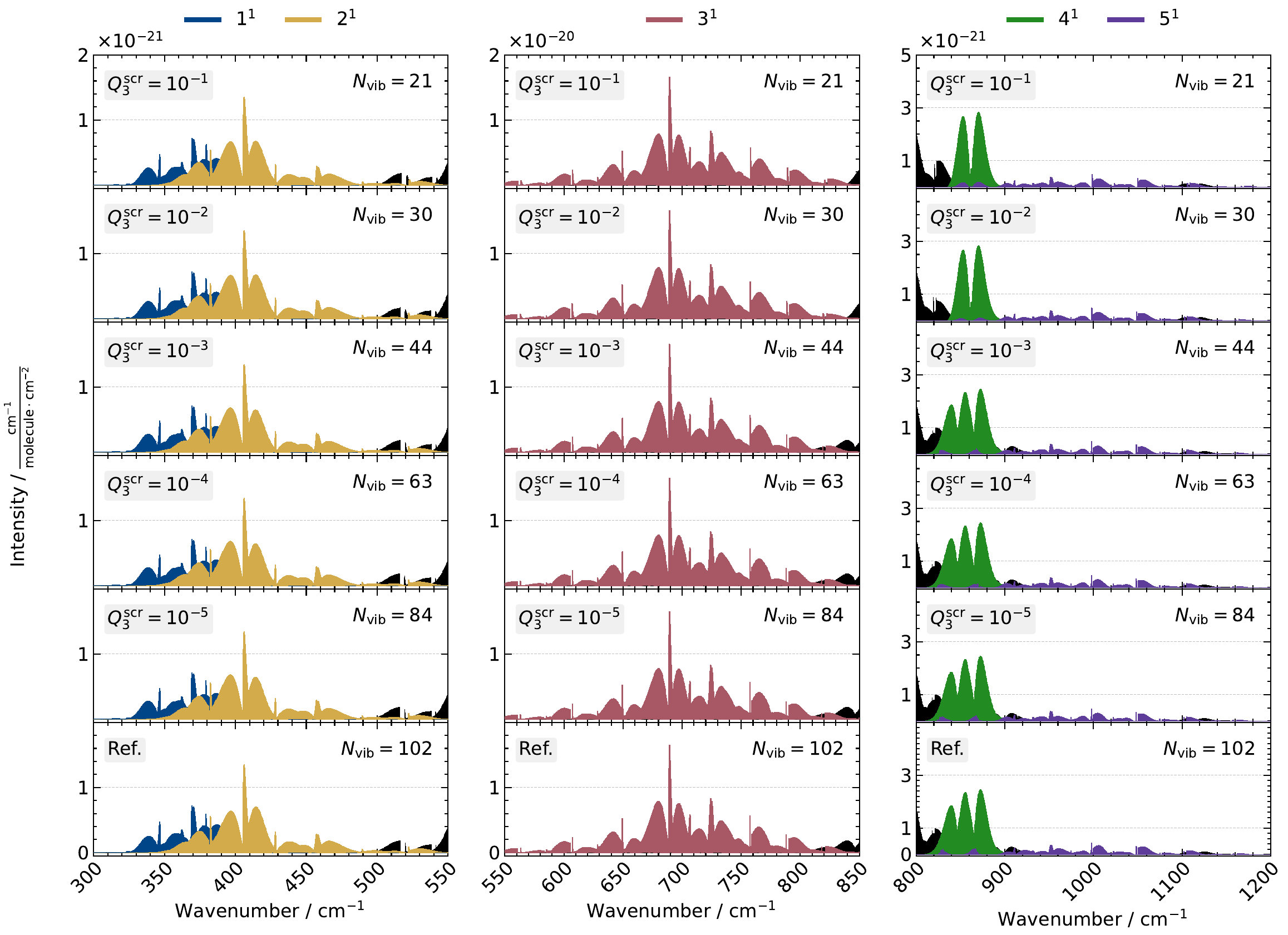}
    \caption{Calculated \ch{H2CCS} RVCI stick spectra at $T=300$~K for the low-frequency spectral region containing the $1^1$--$5^1$ fundamentals. The complete spectra are shown in black, with transitions whose upper states are assigned to the respective target fundamentals highlighted in color. Results are shown for screening thresholds $Q_3^{\rm scr}=10^{-1}$--$10^{-5}$~cm$^{-1}$ and the 102-state reference calculation.}
    \label{fig:h2ccs_spectra}
\end{figure*}

Figure~\ref{fig:h2ccs_spectra} shows the calculated RVCI stick spectra of \ch{H2CCS} for the low-frequency spectral region considered in the main text. The spectra obtained with the pruned vibrational bases for all investigated screening thresholds are compared with the 102-state reference calculation.

\bibliography{literature}

\clearpage

